\documentclass[runningheads,dvipsnames]{llncs}

\usepackage{soul}
\usepackage{url}
\usepackage[hidelinks]{hyperref}
\usepackage[utf8]{inputenc}
\usepackage[T1]{fontenc}
\usepackage[small]{caption}
\usepackage{graphicx}
\usepackage{amsfonts}
\usepackage{amsmath}
\usepackage{pifont}
\usepackage{booktabs}
\usepackage{algorithm}
\usepackage[noend]{algpseudocode}
\usepackage{xspace}
\usepackage{tikz}
\usepackage{pgfplots}
\usepackage{pgfplotstable}
\usepackage{xcolor}
\usepackage{colortbl}
\usepackage{tabularx}
\pgfplotsset{compat=1.14}
\usetikzlibrary{fit,positioning,calc,shapes,backgrounds,external}
\usetikzlibrary{matrix}
\usepgfplotslibrary{groupplots}
\usepackage{hyphenat}
\usepackage{balance}
\usepackage{multirow}
\usepackage{hyperref}
\usepackage{nicefrac}
\usepackage{enumitem}
\usepackage{subcaption}
\usepackage{wrapfig}
\newcommand{\spara}[1]{\smallskip\noindent{\bf #1}}

\newcommand{\para}[1]{\noindent{\bf #1}}

\algnewcommand\Params{\item[\textbf{Params:}]}

\definecolor{cycle1}{RGB}{228, 26, 28}
\definecolor{cycle2}{RGB}{55, 126, 184}
\definecolor{cycle3}{RGB}{77, 175, 74}
\definecolor{cycle4}{RGB}{152, 78, 163}
\definecolor{cycle5}{RGB}{255, 127, 0}
\definecolor{cycle6}{RGB}{153, 153, 153}%{255, 255, 51}
\definecolor{cycle7}{RGB}{166, 86, 40}
\definecolor{cycle8}{RGB}{247, 129, 191}

\definecolor{eyecancerpink}{RGB}{252, 15, 192}

\newcommand{\thiswork}{\textrm{GRASP}\xspace} % easier to catch in all-capitals
\newcommand*{\NP}{$\mathbf{NP}$\xspace}

\newcommand*{\NPcomplete}{$\mathbf{NP}$-complete\xspace}

\newcommand*{\bigO}{\mathcal{O}}

\newcommand{\cmark}{\textcolor{cycle3}{\ding{52}}} %Davide likes 'em colored
\newcommand{\xmark}{\textcolor{cycle1}{\ding{56}}}

\newcommand{\lapl}{\mathcal{L}}

\newcommand{\eye}{I}
\newcommand{\off}{\textsf{off}}

\newcommand{\diag}{\textsf{diag}}

\newcommand{\reals}{\mathbb{R}}

\newcommand{\dm}[1]{#1}
\newcommand{\dmw}[1]{#1}
\newcommand{\jh}[1]{#1}
\newcommand{\pk}[1]{#1}

\newcommand{\pki}[1]{#1} %pk ijcai2021

\newcommand{\pkw}[1]{#1} %pk www2021
\newcommand{\dmt}[1]{#1}
\newcommand*{\Gr}{G}
\newcommand{\V}{V}
\newcommand{\E}{E}

\newcommand{\alfun}{\tau}
\newcommand{\fmap}{T_\mathcal{F}}

\title{\thiswork: Graph Alignment through Spectral Signatures} % \thiswork looks ugly % ok with rm

\author{Judith Hermanns\inst{1} \and
Anton Tsitsulin\inst{2} \and
Marina Munkhoeva\inst{3}\and
Alex Bronstein\inst{4} \and
Davide Mottin\inst{1}\and
Panagiotis Karras\inst{1}
}
\authorrunning{F. Author et al.}
\institute{Aarhus University\\ 
\email{\{davide,judith,panos\}@cs.au.dk}
\and
University of Bonn\\
\email{tsitsulin@bit.uni-bonn.de}
\and
Skoltech\\
\email{marina.munkhoeva@skolkovotech.ru}
\and
Technion\\
\email{bron@cs.technion.ac.il}
	}
\begin{document}

\maketitle

%!TEX root=../main.tex
\begin{abstract}
\pki{What is the best way to match the nodes of two graphs? This \emph{graph alignment} problem generalizes graph isomorphism and arises in applications from social network analysis to bioinformatics.} \pki{Some} solutions assume that auxiliary information on known matches or node or edge attributes \pki{is} available, or utilize arbitrary graph features. Such methods fare poorly in the pure form of the problem, in which only graph structures are given. \pki{Other proposals translate the problem to one of aligning node embeddings, yet, by doing so, provide only a single-scale view of the graph.}

\pki{In this paper, we transfer} \pkw{the shape-analysis concept of functional maps from the continuous to the discrete case, and treat the graph alignment} problem as a special case of the problem of finding a mapping between functions on graphs. \pki{We present \thiswork, a method that first establishes a correspondence between functions derived from Laplacian matrix eigenvectors, which capture multiscale structural characteristics, and then exploits this correspondence to align nodes.} Our experimental study\pkw{, featuring noise levels higher than anything used in previous studies,} shows that \thiswork outperforms state-of-the-art methods for graph alignment across noise levels and graph types.
% already mentioned
%Once this mapping is found, we can treat node-to-node matching as a special case.
\end{abstract}
%!TEX root=../main.tex 
\section{Introduction}\label{sec:introduction}

%Graphs naturally arise as a way to elegantly model dependencies, connections and interactions of arbitrary complexity among arbitrary entities. %In nature, protein networks and enzymes can be interpreted as graphs, comprising valuable information about fundamental biological processes. In recent years, social networks established as an influential component in social and political discourses. 
%They are used across very many domains, including social networks, molecular chemistry, protein networks, knowledge graphs, citation networks, communication networks or the web itself. These and many other graphs contain an extensive amount of information. This information is in most cases hidden on a structural level: The connectivity patterns of nodes form structures, which are characteristic for, e.g., bot networks in social media or for the functioning of proteins. Identifying and re-identifying those structures from small neighborhoods to the full graph allows important insight when analyzing large-scale graphs and large graph repositories: We can, for instance, obtain a deeper understanding of the flow of information in social networks or identify proteins of similar function in different species. However, performing an algorithmic analysis of graphs is a challenging task due to the complex and unordered nature of graphs and also their size and number, which in real-world scenarios often can be excessive.

% opening: what graphs are
\pk{Graphs model relationships between entities in several domains, e.g., social networks, protein interaction networks, email communication or chemical molecules.
The structure of such graphs captures rich information on how people are connected, how molecules function, or how proteins interact.}

% other side of the matter % graphs are not disordered, they are complex; a problem cannot appear to be basic
\pk{At the same time, the expressive nature of graphs also implies complexity, which renders some fundamental problems hard. 
For instance, the \emph{graph isomorphism} problem, which is to determine whether two graphs share the same structure is neither known to be polynomially solvable nor \NPcomplete, and has been used to define the $\mathbf{GI}$ complexity class~\cite{kobler12}. 
Problems that generalize graph isomorphism occur frequently in the field of graph analytics. 
One of those is the \NPcomplete \emph{subgraph isomorphism} problem; another is \emph{graph alignment}, which aims to find the best (exact or inexact) matching among the nodes of a pair of graphs; a solution to this problem is sine qua non in tasks such as identifying users in different social networks~\cite{kazemi2015growing}, matching objects in images by establishing feature correspondences ~%\cite{schellewald2005probabilistic}, 
and \dm{comprehending protein response in the body~\cite{klau2009new}.}

In case additional background information is available, such as node and edge attributes in the two graphs to be aligned, or existing valid \emph{seed} matches, \dm{then the problem \pk{is solvable} via supervised methods~\cite{liu2016aligning,chu2019cross}.} However, in case only graph structures are given, then the problem \dm{of aligning two graphs by matching structures,} is at least as hard as graph isomorphism \dm{even in its approximate version~\cite{abdulrahim1998parallel}.}}

%!TEX root=../ijcai21.tex

\begin{figure}[!t]
\centering
\subcaptionbox{Karate club; Red edges removed.}[.45\columnwidth]{
\centering
\resizebox{.48\columnwidth}{!}{\begin{tikzpicture}
\foreach \nodename/\x/\y/\nc in {
0/1.3404203721530208/-0.20792319009521731, 1/0.660612797683139/-0.5293529379816081, 2/-0.0842929565344319/-0.19435702604686805, 3/0.9887056104542613/-0.6556772663132004, 4/2.388726281533438/0.4359886477065871, 5/3.03864089819031/-0.0815380112738996, 6/2.960969662420727/0.2471675252716638, 7/0.6686122910347339/-0.8852599447509695, 8/-0.35755266045473977/-0.02179343136352086, 9/-1.106635392656473/-0.7165162549717286, 10/2.4966532693182515/0.042620850132054605, 11/2.649380097135513/-0.7839021261548074, 12/1.7682909761217154/-1.170201089287862, 13/0.32232889198239634/-0.12036557215242003, 14/-2.537820929766365/-0.1595232890784888, 15/-2.017977218523939/-0.5868883852966252, 16/4.0/0.20689695091332871, 17/1.5351649410774988/-0.9284844629186799, 18/-2.361501070585379/-0.4157678419029558, 19/0.010928809575181745/-0.48917814673461135, 20/-2.6366053895646013/0.12295206842632095, 21/1.824084503111609/-0.5719597768990051, 22/-2.4105548253974125/0.37339897206584516, 23/-1.5384781962852379/0.9162452957787691, 24/-0.4005990099812155/1.4907649184596932, 25/-0.9487157271880506/1.5137087504718347, 26/-2.6631474476122494/0.8212188725418305, 27/-0.7941114551056272/0.7930274430468092, 28/-0.6848060249576489/0.36407613033716624, 29/-2.1118981360210554/0.7252382670203417, 30/-0.7361185161547416/-0.3791804505545907, 31/-0.31651894133461983/0.6432253802633714, 32/-1.5787398457807174/0.08778721720777753, 33/-1.3674456578872773/0.11355191413366289
}
{
\node (\nodename) at (\x,\y) [shape=circle,inner sep=3pt,draw,thick,fill=cycle2] {}; % cycle2 or cycle5
}
\begin{pgfonlayer}{background}
\path
\foreach \startnode/\endnode in {
0/1, 0/2, 0/3, 0/4, 0/5, 0/6, 0/7, 0/8, 0/10, 0/11, 0/12, 0/13, 0/17, 0/19, 0/21, 0/31, 1/17, 1/2, 1/3, 1/21, 1/19, 1/7, 1/13, 1/30, 2/3, 2/32, 2/7, 2/8, 2/9, 2/27, 2/28, 2/13, 3/7, 3/12, 3/13, 4/10, 4/6, 5/16, 5/10, 5/6, 6/16, 8/32, 8/30, 8/33, 9/33, 13/33, 14/32, 14/33, 15/32, 15/33, 18/32, 18/33, 19/33, 20/32, 20/33, 22/32, 22/33, 23/32, 23/25, 23/27, 23/29, 23/33, 24/25, 24/27, 24/31, 25/31, 26/33, 26/29, 27/33, 28/33, 28/31, 29/32, 29/33, 30/33, 30/32, 31/32, 31/33, 32/33}
{
(\startnode) edge[-,thick,draw opacity=0.75] node {} (\endnode)
};
\path
\foreach \startnode/\endnode in {
0/1, 0/8, 0/13, 1/17, 3/12, 3/13, 14/33, 18/32, 23/33, 24/27, 25/31, 31/32, 32/33}
{
(\startnode) edge[-,ultra thick, draw=cycle1] node {} (\endnode)
};
\end{pgfonlayer}
\end{tikzpicture}}
}\hspace{2mm}
\subcaptionbox{Alignment by \thiswork{} (top) and REGAL (bottom).}[.45\columnwidth]{
\centering
\resizebox{.3\columnwidth}{!}{\begin{tikzpicture}
\foreach \nodename/\x/\y in {
0/1.3404203721530208/-0.20792319009521731, 1/0.660612797683139/-0.5293529379816081, 2/-0.0842929565344319/-0.19435702604686805, 3/0.9887056104542613/-0.6556772663132004, 5/3.03864089819031/-0.0815380112738996, 6/2.960969662420727/0.2471675252716638, 7/0.6686122910347339/-0.8852599447509695, 9/-1.106635392656473/-0.7165162549717286, 13/0.32232889198239634/-0.12036557215242003, 16/4.0/0.20689695091332871, 18/-2.361501070585379/-0.4157678419029558, 19/0.010928809575181745/-0.48917814673461135, 20/-2.6366053895646013/0.12295206842632095, 22/-2.4105548253974125/0.37339897206584516, 23/-1.5384781962852379/0.9162452957787691, 24/-0.4005990099812155/1.4907649184596932, 25/-0.9487157271880506/1.5137087504718347, 26/-2.6631474476122494/0.8212188725418305, 27/-0.7941114551056272/0.7930274430468092, 28/-0.6848060249576489/0.36407613033716624, 29/-2.1118981360210554/0.7252382670203417, 31/-0.31651894133461983/0.6432253802633714, 32/-1.5787398457807174/0.08778721720777753, 33/-1.3674456578872773/0.11355191413366289
}
{
\node (\nodename) at (\x,\y) [shape=circle,inner sep=3pt,draw,thick,fill=cycle3] {}; % good nodes
}
\foreach \nodename/\x/\y in {
4/2.388726281533438/0.4359886477065871, 8/-0.35755266045473977/-0.02179343136352086, 10/2.4966532693182515/0.042620850132054605, 11/2.649380097135513/-0.7839021261548074, 12/1.7682909761217154/-1.170201089287862, 14/-2.537820929766365/-0.1595232890784888, 15/-2.017977218523939/-0.5868883852966252, 17/1.5351649410774988/-0.9284844629186799, 21/1.824084503111609/-0.5719597768990051, 30/-0.7361185161547416/-0.3791804505545907
}
{
\node (\nodename) at (\x,\y) [shape=circle,inner sep=3pt,draw,thick,fill=cycle1] {}; % bad nodes
}
\begin{pgfonlayer}{background}
\path
\foreach \startnode/\endnode in {
2/0, 2/1, 2/3, 2/7, 2/8, 2/9, 2/13, 2/27, 2/28, 2/32, 0/3, 0/4, 0/5, 0/6, 0/7, 0/10, 0/11, 0/12, 0/17, 0/19, 0/21, 0/31, 3/1, 3/7, 4/6, 4/10, 5/6, 5/10, 5/16, 6/16, 7/1, 19/1, 19/33, 21/1, 31/24, 31/28, 31/33, 1/13, 1/30, 13/33, 30/8, 30/32, 30/33, 8/32, 8/33, 9/33, 27/23, 27/33, 28/33, 32/14, 32/15, 32/20, 32/22, 32/23, 32/29, 33/15, 33/18, 33/20, 33/22, 33/26, 33/29, 25/23, 25/24, 23/29, 29/26}
{
(\startnode) edge[-,thick,draw opacity=0.75] node {} (\endnode)
};
\end{pgfonlayer}
\end{tikzpicture}}
\resizebox{.3\columnwidth}{!}{\begin{tikzpicture}
\foreach \nodename/\x/\y in {
4/2.388726281533438/0.4359886477065871, 10/2.4966532693182515/0.042620850132054605, 11/2.649380097135513/-0.7839021261548074, 16/4.0/0.20689695091332871, 18/-2.361501070585379/-0.4157678419029558, 23/-1.5384781962852379/0.9162452957787691, 24/-0.4005990099812155/1.4907649184596932, 25/-0.9487157271880506/1.5137087504718347, 31/-0.31651894133461983/0.6432253802633714
}
{
\node (\nodename) at (\x,\y) [shape=circle,inner sep=3pt,draw,thick,fill=cycle3] {}; % cycle2 or cycle5
}
\foreach \nodename/\x/\y in {
0/1.3404203721530208/-0.20792319009521731, 1/0.660612797683139/-0.5293529379816081, 2/-0.0842929565344319/-0.19435702604686805, 3/0.9887056104542613/-0.6556772663132004, 5/3.03864089819031/-0.0815380112738996, 6/2.960969662420727/0.2471675252716638, 7/0.6686122910347339/-0.8852599447509695, 8/-0.35755266045473977/-0.02179343136352086, 9/-1.106635392656473/-0.7165162549717286, 12/1.7682909761217154/-1.170201089287862, 13/0.32232889198239634/-0.12036557215242003, 14/-2.537820929766365/-0.1595232890784888, 15/-2.017977218523939/-0.5868883852966252, 17/1.5351649410774988/-0.9284844629186799, 19/0.010928809575181745/-0.48917814673461135, 20/-2.6366053895646013/0.12295206842632095, 21/1.824084503111609/-0.5719597768990051, 22/-2.4105548253974125/0.37339897206584516, 26/-2.6631474476122494/0.8212188725418305, 27/-0.7941114551056272/0.7930274430468092, 28/-0.6848060249576489/0.36407613033716624, 29/-2.1118981360210554/0.7252382670203417, 30/-0.7361185161547416/-0.3791804505545907, 32/-1.5787398457807174/0.08778721720777753, 33/-1.3674456578872773/0.11355191413366289
}
{
\node (\nodename) at (\x,\y) [shape=circle,inner sep=3pt,draw,thick,fill=cycle1] {}; % cycle2 or cycle5
}
\begin{pgfonlayer}{background}
\path
\foreach \startnode/\endnode in {
2/0, 2/1, 2/3, 2/7, 2/8, 2/9, 2/13, 2/27, 2/28, 2/32, 0/3, 0/4, 0/5, 0/6, 0/7, 0/10, 0/11, 0/12, 0/17, 0/19, 0/21, 0/31, 3/1, 3/7, 4/6, 4/10, 5/6, 5/10, 5/16, 6/16, 7/1, 19/1, 19/33, 21/1, 31/24, 31/28, 31/33, 1/13, 1/30, 13/33, 30/8, 30/32, 30/33, 8/32, 8/33, 9/33, 27/23, 27/33, 28/33, 32/14, 32/15, 32/20, 32/22, 32/23, 32/29, 33/15, 33/18, 33/20, 33/22, 33/26, 33/29, 25/23, 25/24, 23/29, 29/26}
{
(\startnode) edge[-,thick,draw opacity=0.75] node {} (\endnode)
};
\end{pgfonlayer}
\end{tikzpicture}}
}
%\vspace*{-3mm}
\caption{With a few removed edges, REGAL~\protect\cite{heimann18regal}, \pki{an} alignment method based on \emph{local} features, fails to correctly align the distorted Karate club graph to the original; \thiswork identifies most of nodes (correctly aligned nodes in green).} \label{fig:firstpage}
\vspace{-5mm}
\end{figure}
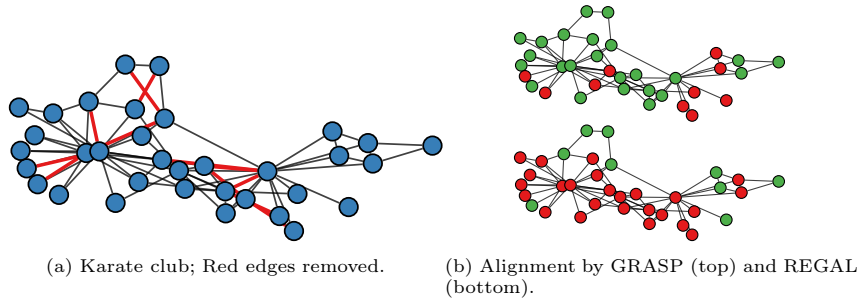

% what is bad about techniques used so far
\pk{Existing approaches to the graph alignment problem are oriented toward using a few \emph{heuristic graph features}, such as landmarks, in order to detect a good alignment~\cite{heimann18regal}, \dm{\emph{exploiting additional information} such as node attributes~\cite{zhang2016final} or bipartite networks~\cite{koutra2013bigalign}, or optimizing objectives based only on \emph{local connections} among nodes~\cite{feizi2019spectral,nassar2018lowrank,liao2009isorankn}.} 
On the other hand, the spectra of \emph{Laplacian matrices} have been successfully employed to devise a similarity measure among graphs~\cite{tsitsulin18kdd}. 
\dm{Laplacian spectra capture important \emph{multiscale} properties, such as local-scale ego-nets and global-scale communities. 
Previous approaches rooted in spectral characteristics decompose large matrices expressing all alignments among edges in two graphs~\cite{feizi2019spectral,nassar2018lowrank,liao2009isorankn} and formulate the solution as finding the leading eigenvector of such matrices. These approaches disregard \pkw{most} eigenvectors and consider only local edge variations. To our knowledge, the spectral properties of \pkw{\emph{Laplacian matrices} have \emph{not yet} been utilized to any significant extent} for an end-to-end graph alignment method.}}
%\cmt{too strong claim, find qualifying features that make this statement valid even when considering other methods, provide information here.}}

\pk{
We propose \dm{\thiswork, short for \textbf{GR}aph \textbf{A}lignment through \textbf{SP}ectral Signatures,} a principled approach towards detecting a good alignment among graphs, grounded on their spectral characteristics, i.e., eigenvalues and eigenvectors of their Laplacian matrices~\cite{chung1997spectral}. We transfer the methodology of matching among shapes based on \emph{corresponding functions}~\cite{DBLP:journals/tog/OvsjanikovBSBG12} to the domain of graphs: we first extract a mapping of node-evaluated functions grounded on the graph's heat kernel, and then apply this mapping to the matching on nodes.} \dm{Figure \ref{fig:firstpage} shows an example alignment of the Karate club with a deteriorated version obtained by removing some edges; \thiswork correctly aligns most of the nodes, while REGAL~\cite{heimann18regal} based on local descriptors fails to do so.}

\pki{In short, we propose \thiswork, a graph alignment method based on spectral graph characteristics and show its effectiveness in recovering real-graph alignments, with higher accuracy and similar runtime as the state of the art.} % what else?

%This paper is organized as follows....
%\begin{itemize}
%    \item Graphs are important because ...
%    \item Graph alignment is important because ...
%    \item Drawback of current approaches
%    \begin{itemize}
%        \item $\min _p|| PAP^T-B||^2$ find P methods
%        \item embedding methods/similarity based methods
%    \end{itemize}
%     \item Rigid node-to-node matching that does not take into account structural information
%        \item Rely on strict constraints for node-to-node matching that do not meet reality
%    \item they all don't do structure
%    \item spectrum of a graph and why it is a good idea regarding structure
%    \item We steal from computational geometry
%    \item Our contribution: ...
%\end{itemize}

%!TEX root=../main.tex
\section{Related Work}\label{sec:related-work}

% Flexibility
% Spectral
% Multiscale - Hierarchical
% Complexity

% the opposite of unrestruicted is restricted.
% DM Jan 19 - We already said that. 
%\pk{The need for accurate graph alignment arises in several disciplines, including bioinformatics~\cite{hashemifar2014hubalign, saraph2014magna, aladaug2013spinal}, computer vision~\cite{schellewald2005probabilistic, cho2010reweighted, zanfir2018deep}, and social network analysis~\cite{kazemi2015growing, yartseva2013performance, pedarsani2011privacy, kazemi2015can}. 
\pk{We discuss related work in two main categories: {\bf restricted alignment}, which requires ground-truth mapping or other additional information, and {\bf unrestricted alignment}, which requires neither supervision nor additional information.} \dm{Table \ref{tbl:relatedwork} gathers together previous works' characteristics.} 
%\cmt{Why are we talking about graph matching here?}
%\cmt{Maybe add something about multi-network alignment}

\subsection{Restricted Alignment}
% sounds better than constrained, especially for something we do not do.

% We explain on the way % but we need to start with an overview.
\pki{\emph{Restricted} methods incorporate non-structural information; a restricted method can be \emph{supervised} or \emph{assisted}.}

\pki{\spara{Supervised methods} exploit pre-aligned pairs of seed nodes to construct a first alignment. \emph{Percolation graph matching} (PGM)~\cite{kazemi2015growing,yartseva2013performance} propagates ground-truth alignments across the network.
%using percolation theory.
%; they are useful in social network analysis and grounded in percolation theory. 
\emph{Representation learning} approaches, such as IONE~\cite{liu2016aligning}, PALE~\cite{man2016predict}, and DeepLink~\cite{zhou2018deeplink}, learn a low-dimensional embedding of the graph nodes and map the node embeddings of one graph to another. A similar method aligns multiple networks at once~\cite{chu2019cross}. \emph{Active network alignment}~\cite{malmi2017active} applies active learning %(i.e., seeks supervision on its own)
to elicit expert guidance on alignments. Overall, supervised methods rely on prior knowledge, which may not be available.}
%, which is the case only in specific domains; 
%such methods do not transfer well to other domains.
% IONE~\cite{liu2016aligning}: Takes into account also node-neighborhood. 
% PALE~\cite{man2016predict}: learn representations with 
% \cite{zhou2018deeplink}: uses anchor nodes, extends PALE with mapping direction and random walks. 

\pk{\spara{Assisted methods} utilize auxiliary information or structural constraints.
%other than seed nodes. 
BigAlign~\cite{koutra2013bigalign} \pki{focuses on} bipartite graphs; 
%; in this case, the task is relatively easier, as bipartite graphs are partitioned into two groups of nodes with no in-group edges;
however, most graphs are not bipartite. FINAL~\cite{zhang2016final} \pki{aligns nodes based on similarity of topology and attributes.}
% formulates an objective function based on the \emph{alignment consistency} principle, according to which, if one pair of nodes is similar to another pair of nodes in terms of topology and attributes, then it should also be similar in terms of alignment.
\pki{\jh{GSANA~\cite{yasar2018iterative}} lets pairwise distances to seed nodes guide the matching.}
%is a method on the border of supervised and assisted methods. Using 
%employs a set of seed nodes, which can be given by the user or precomputed, to calculate pairwise distances used as 2D-coordinates for matching.
%the method calculates the pairwise distances to each pair of seeds, uses these as 2D-coordinates and uses these for matching. If \pkw{seeds} are not given, a matrix with prior matching information is used.
\pki{Another variant matches} \emph{weigthed} matrices using their spectra~\cite{umeyama1988}; that is inapplicable to the unweighted case.} Overall, \pki{restricted} methods cannot handle cases where the only given information is graph structure.
% pk added Umeyama 20oct2020

% this apparently belongs here.
% Unsupervised: Constraints (bipartite)~\cite{koutra2013bigalign} or attributes 
%Other things that are often used in addition to structure are node or edge attributes. \cite{heimann18regal, zhang2016final, DBLP:conf/pakdd/HeimannLPCK18}. FINAL \cite{zhang2016final} proposes to use attribute similarity together with topology consistency in order to optimize a quadratic objective function. Here topology consistency denotes that if two pairs of nodes are matched across graphs, the relationship between the nodes within their graphs should be consistent across graphs.

%!TEX root = ../ijcai21.tex
%%% Hash-align???

\begin{table}[t!]
\setlength{\aboverulesep}{0pt}
\setlength{\belowrulesep}{0pt}

% \begin{center}
{\small
%\scriptsize
\newcolumntype{C}{>{\centering\arraybackslash}X}
\newcolumntype{H}{>{\setbox0=\hbox\bgroup}c<{\egroup}@{}}
%\begin{tabular}{l*{6}{>{\centering\arraybackslash}m{1.25cm}}}
\begin{tabularx}{\linewidth}{p{1.5cm}CCCCCC} %\columnwidth
%\toprule
% \multicolumn{1}{C}{} & \multicolumn{4}{c}{\textbf{Properties}} \\%& \multicolumn{2}{c}{\textbf{Complexity}} \\
%\cmidrule(lr){2-5}\cmidrule(lr){6-7}
\bf{Method} & \rotatebox{35}{Unsuperv.} & \rotatebox{35}{Spectral} & \rotatebox{35}{Plain} & \rotatebox{35}{Flexible} & \rotatebox{35}{\parbox[t]{1.5cm}{Offline \\ Precomp.}} & \rotatebox{35}{Multiscale}\\ %& Alignment\\ 
\midrule
		    % & \cmark{} & \xmark{}   & \xmark{}  & $\bigO(1)$            & \NPcomplete{}	\\ % Min-complexity for exact? 
		    % & \xmark{} & \xmark{}   & \xmark{}  & $\bigO(1)$            & $\bigO(m)$	\\
Supervised
%~\cite{liu2016aligning,chu2019cross,man2016predict,zhou2018deeplink}	
& 	\xmark & \xmark    & \cmark{} & \xmark    	& \xmark  & \xmark{}	\\%& $\bigO(1)$            & $\bigO(n^2)$	\\
BigAlign%~\cite{koutra2013bigalign}	
& 	\cmark & \xmark{} & \cmark & \xmark{}   & \xmark{}  & \xmark \\%& $\bigO(1)$            & $\bigO(n^3)$	\\
FINAL%~\cite{zhang2016final}		 
&  \cmark  & \xmark{} & \xmark & \xmark{}   & \xmark{} & \xmark{}  \\%& $\bigO(k m + k^2 n)$  & $\bigO(k^3)$	\\
\midrule
IsoRank%~\cite{singh2008global}/IsoRankN~\cite{liao2009isorankn} 
&  \cmark  & \xmark{} & \cmark{}  & \cmark & \cmark & \xmark  \\%& $\bigO(k m + k^2 n)$  & $\bigO(k^3)$	\\
LaplMatch
&  \cmark  & \cmark{} & \cmark{}  & \xmark & \cmark & \xmark  \\%& $\bigO(k m + k^2 n)$
LREA%~\cite{nassar2018lowrank}		
&   \cmark  & \cmark{} & \cmark{} 	& \xmark{} & \xmark & \xmark	\\%& $\bigO(n \log n)$ 	& $\bigO(1)$	\\
REGAL%~\cite{heimann18regal}		
&  \cmark  & \xmark{} & \cmark{}  & \cmark	& \cmark{} & \xmark \\%& $\bigO(n^2)$ 			& $\bigO(1)$	\\
\midrule
\rowcolor{cycle2!10}
\textbf{\thiswork{}} & \cmark & \cmark{} & \cmark{} &  \cmark & \cmark{}  & \cmark\\%& $\bigO(k m + k^2 n)$ & $\bigO(1)$\\
%\bottomrule
\end{tabularx}
}
% \end{center}
\caption{Related work in terms of \pkw{present} (\cmark) and \pkw{absent} (\xmark) properties. \emph{Supervised} methods
~\protect\cite{liu2016aligning,chu2019cross,man2016predict,zhou2018deeplink} require aligned nodes as input. \emph{Spectral} methods~\protect\cite{nassar2018lowrank} use spectral properties of alignment matrices. FINAL~\protect\cite{zhang2016final} does not work on \emph{plain} graph structures as it requires node attributes. REGAL~\protect\cite{heimann18regal} and IsoRank~\protect\cite{liao2009isorankn,singh2008global} are \emph{flexible} in allowing different algorithms for alignment (e.g.\ bipartite matching, nearest neighbors). \pkw{\pki{Among unrestricted methods (rows~4--7), LREA~\protect\cite{nassar2018lowrank} cannot} benefit from offline precomputation
% prior to alignment % definition of "precomputation"
(results in Figure~\ref{fig:time}).}
\pkw{\thiswork explicitly captures \emph{multiscale} properties through the heat kernel.} } 
% Partially replying to rev. 4
\label{tbl:relatedwork}
\vspace{-4mm}
\end{table}

\subsection{Unrestricted Alignment}

% Removed to avoid redundancy and achieve compactness % we need opening!
\pki{\emph{Unrestricted} methods require neither prior knowledge of ground-truth pairs nor other information on the input graph.}

\para{Integer-programming methods.} Klau~\cite{klau2009new} presents a Lagrangian relaxation for the integer programming problem posed by network alignment; the resulting algorithm is polynomial, yet still impracticable for large networks.

\spara{Embedding-based methods.} REGAL~\cite{heimann18regal} constructs node embeddings based on the connectivity structure and node attributes, and uses the similarity between these features for node alignment; we classify REGAL as an unrestricted method since it can work without attributes. \pki{CONE-Align~\cite{chen2020cone} realigns node representations, without prejudice to the representation used.} % remark on employed our method goes to Conclusions.

\spara{Matrix decomposition methods.} IsoRank~\cite{singh2008global} aligns multiple protein-protein interaction networks aiming to maximize the overall quality across all input networks; it constructs an eigenvalue problem for every pair of input networks and extracts a global alignment across a set of networks by a $k$-partite matching; it uses structural properties (PageRank), but also relies on a similarity measure between nodes which in a biology-specific case builds on the similarity of the proteins; it is improved with greedy approaches~\cite{kollias2013fast}. Another improvement on IsoRank, IsoRankN~\cite{liao2009isorankn}, performs spectral clustering on the induced graph of pairwise alignment scores; as it is based on spectral methods, IsoRankN is claimed to be both error-tolerant and computationally efficient.
% this is also a spectral method!
EigenAlign~\cite{feizi2019spectral} formulates the problem as a Quadratic Assignment Problem that considers both matches and mismatches and solves it by \emph{spectral decomposition} of matrices. Building thereupon, Low-Rank EigenAlign~\cite{nassar2018lowrank} solves a maximum weight bipartite matching problem on a \emph{low-rank} version of a node-similarity matrix, hence requires memory linear in the size of the graphs. However, EigenAlign variants use the first eigenvector of a joint adjacency matrix between the two graphs to be aligned, rather than the eigenvectors of graph Laplacians, which provides richer information. 
%A projected power iteration version of EigenAlign, Projected Power Alignments (PPA)~\cite{onaran2017projected}, improves recovery rates.
%, CrossNA~\cite{chu2019cross} % what is this for here?

\spara{Belief propagation methods.} NetAlign \cite{bayati2013message} \pkw{solves a \emph{sparse} graph alignment variant by} message passing.

\subsection{Shape Matching} 

\pk{Our work is inspired by shape matching methods that employ spectral properties~\cite{litany17,DBLP:journals/tog/OvsjanikovBSBG12,kovnatsky13coupled}. Functional maps~\cite{DBLP:journals/tog/OvsjanikovBSBG12} generalize the matching of points to the matching of \emph{corresponding functions} among shapes, by revealing a common decomposition of such functions using the eigenvectors of the Laplace\hyp{}Beltrami operator; the graph equivalent of that operator is a graph's Laplacian matrix. Extensions of this methods match non-isometric shapes by aligning their Laplace\hyp{}Beltrami operators' eigenbases~\cite{kovnatsky13coupled}, and match a part of a shape to another full shape
%~\cite{rodola17}; such partial matching can be done fully 
in the spectral domain~\cite{litany17} without requiring spatially modeling the part of a shape.}

\subsection{Spectral Methods} 

\pki{Graph spectra~\cite{chung1997spectral} facilitate problem-solving in graph analysis, image partitioning, graph search, and machine learning~\cite{shi2000normalized}. 
% Redunant: we describe this in detail in Section 4.8
%The eigenvectors of the Laplacian of a point cloud graph converge to the eigenfunctions of the Laplace\hyp{}Beltrami operator on the underlying %Riemannian manifold, which justifies transferring Laplace\hyp{}Beltrami operator-based methods from computational geometry to similar problems in graph analysis. 
NetLSD~\cite{tsitsulin18kdd} uses Laplacian spectral signatures to detect graph similarity\pki{, \emph{but not to align graphs},} in a multi-scale fashion. 
\pki{LaplMatch~\cite{knossow2009inexact} derives a permutation matrix for shape matching from Laplacian eigenvectors, without considering multiscale properties.}
%Graph convolutional networks also utilize graph spectra~\cite{defferrard2016convolutional} \dm{to learn filters on the eigenvectors}.
While calculating a graph's spectrum is computationally challenging, recent work proposes an approximation via spectral moments estimated through random walks~\cite{DBLP:conf/kdd/Cohen-SteinerKS18}.}
% Seems out of place here -- We already described faster method. 
%\pkw{Our work employs graph spectra, yet can rely on fast methods for diagonally dominant matrices~\cite{koutis2015faster}.}
%!TEX root=../main.tex
\section{Background and Problem}\label{sec:problem}

\spara{Graph Alignment.} \dmw{Consider two undirected graphs, $\Gr_1 = (V_1, E_1)$ and $\Gr_2 = (V_2, E_2)$, where $V_*$ are node sets, $E_* \subseteq V_* \times V_*$ are edges, \pki{and\footnote{\pki{Solutions to the problem of aligning graphs with unequal numbers of nodes can rest on solutions to this basic problem form.}}} $|V_1|=|V_2|=n$. A graph's \emph{adjacency matrix} $A \in \{0,1\}^{n\times n}$ is a binary matrix where $A_{ij}=1$ if there is an edge between nodes $i$ and $j$ and $A_{ij}=0$ otherwise.}

\vspace{-1mm}
\begin{definition}\label{def:alignment}
Given two graphs $\Gr_1=(V_1,E_1)$ and $\Gr_2=(V_2,E_2)$, a graph alignment $R:V_1 \rightarrow V_2$ is an \emph{injective} function that maps nodes of $\Gr_1$ to nodes of $\Gr_2$.
\end{definition}
\vspace{-1mm}

The graph alignment problem is to find such a function, which, expressed as a permutation matrix $P$, minimizes the difference $\|PA_1P^\top-A_2\|^2$. In case of isomorphic graphs, there exists a $P$ such that $PA_1P^\top = A_2$, i.e., aligns the two graphs exactly. We are interested in the general, unrestricted problem case, in which there are no additional constraints on node attributes or matches known in advance. The problem is hard and not known to be in \NP.%~\cite{cook71}. 

\dmw{We may express graph alignment in terms of a ground truth function $\alfun: V_1 \rightarrow V_2$ that returns the correct alignment between the nodes $V_1$ in $\Gr_1$ and the nodes $V_2$ in $\Gr_2$. In the case of isomorphic graphs, \pkw{this} ground truth function $\alfun$ is a bijection \pkw{that} admits an inverse mapping $\pkw{\alfun^{-1}: V_2 \rightarrow V_1}$. \pkw{The composition of} the indicator function~$\delta_i: \V_1 \rightarrow \{0,1\}$ 
% \[\delta_i(j)=\begin{cases}
% 1 & \text{if } j=i\\
% 0& \text{else}  
% \end{cases}\label{eq:delta}\]
\pkw{with}~$\alfun^{-1}$, $\delta_i\circ \alfun^{-1}:\pkw{ V_2 \rightarrow \{0,1\}}$ \pkw{expresses} the complete isomorphism among the two graphs, returning~$1$ if node~$\pkw{u \in \V_2}$ maps to node~\pkw{$i \in \V_1$}, \pkw{$0$ otherwise}. By generalization, the composition $g_i = f_i \circ \tau^{-1}$ maps functions in $\Gr_2$ to functions in~$\Gr_1$ for any family of real-valued functions $f_1, ..., f_q, f_i: \V_1 \rightarrow \reals$ and $g_1, ..., g_q, g_i: \V_2 \rightarrow \reals$ that associate a real value to each node in~$\Gr_1$ and~$\Gr_2$. 
%Such a composition defines an operator $\tau_\mathcal{F}:(V_1 \times \reals) \rightarrow (V_2 \times \reals)$ from functions in $V_$
This transformation among functions is called a \emph{functional representation} of the mapping $\tau$. In effect, finding an alignment among the nodes of two graphs corresponds to finding an alignment among functions on those nodes. We \pkw{use} such \emph{functional alignments} \pk{as a shortcut} to \emph{node alignments}. \pkw{To get there, we extend the concept of a functional map~\cite{DBLP:journals/tog/OvsjanikovBSBG12} \pki{from the continuous to the discrete case}.}
% pk edited 19oct2020 condensed 20jan2021

\spara{Functional maps.} The operator $\fmap:(V_1 \times \reals) {\rightarrow} (V_2 \times \reals)$ maps functions $f$ on the nodes in $\Gr_1$ to functions $g$ on the nodes in $\Gr_2$, i.e. $\fmap(f){=}f{\circ}\tau^{-1}{=}g$. This operator is linear in the function space\pkw{, i.e.,} $\fmap(c_1f_1 + c_2f_2)=(c_1f_1 + c_2f_2)\circ\alfun^{-1} = c_1f_1 \circ \alfun^{-1} + c_2f_2 \circ \alfun^{-1} = c_1\fmap(f_1) + c_2\fmap(f_2)$.
%As such, the $\fmap$ operator is amenable to transformations through linear algebra.
%\smallskip
%\noindent %We now define the concept of \emph{functional alignment}.
\pk{In addition,} let $\phi_1,...,\phi_n$ and $\psi_1, ..., \psi_n$ denote orthogonal bases for the space of functions on $\Gr_1$'s nodes, $V_1 \times \reals$, and that on $\Gr_2$'s nodes, $V_2 \times \reals$, respectively. Since \pk{those functions produce} $n$-dimensional vectors, we can represent \pk{them} as linear combinations of \pk{their} basis vectors, $f=\sum_{i=1}^{n} a_i\phi_i$ and $g=\sum_{j=1}^{n} b_j\psi_j$. Then, by the linearity of $\fmap$,
\begin{equation*}
\resizebox{\linewidth}{!}{$
\displaystyle\fmap(f) {=} \fmap\left(\sum_{i=1}^{n} a_i\phi_i\right) {=} \sum_{i=1}^{n}a_i\fmap(\phi_i) {=} \sum_{i=1}^{n}a_i\sum_{j=1}^{n}c_{ij}\psi_j {=} \sum_{j=1}^{n} b_j\psi_j
$}
\end{equation*}

where $\fmap(\phi_i) = \sum_{j=1}^{n}c_{ij}\psi_j$. It follows that each coefficient $b_j$ is the dot-product $\sum_{i=1}^{n}a_ic_{ij}$ between the coefficients $(a_1, ...., a_n)$ of functions in $\Gr_1$ and the coefficients $(c_{1j}, ...., c_{nj})$ of the operator $\fmap$. In conclusion, \pk{in order to align} real-valued functions on the nodes of two graphs, \pk{we need to} find a \emph{mapping matrix}  $C \in \reals^{n\times n}$ of coefficients among those functions;
% DM: Moved from below
\pk{such a mapping matrix~$C$ maps functions from~$\Gr_1$ to~$\Gr_2$, even when the ground\hyp{}truth mapping~$\tau$ is unknown. In a nutshell, \thiswork obtains such a mapping matrix $C$ for a \pk{well-chosen} function and applies that~$C$ to mapping the indicator function~$\delta$ from~$\Gr_1$ to~$\Gr_2$, thereby constructing a node alignment.}
%In Section~\ref{sec:solution}, we expand this idea to non-isomorphic graphs.
\pk{The main question we need to answer is what orthogonal basis and functions we should use to construct our mapping matrix $C$.}
%At this point, several questions arise: (1) What is a suitable basis to express functions on graphs? (2) what sort of functions can be used for functional alignment? (3) How do we compute $C$ from the functions? (4) How do we return a node-to-node alignment from a functional alignment?
The next section answers this question and builds on that answer to devise a solution.}
%based on spectral graph theory and linear algebra.
% condensed 20jan2021

% DM: Old text 
% A popular way to address graph alignment is to match nodes among the two graphs guided by appropriate node representations. The challenge is then to (i) find such an appropriate node representation that captures graph structure at both local and large scale, going beyond mere node degree or ego-net structure, including neighborhood connectivity and community structure, and (ii) match nodes having similar characteristics on different levels, from their relationship to the whole graph over community membership to node degrees and neighborhoods. 
% \dm{A graph representation that captures such properties is the eigenvectors and eigenvalues of the graph Laplacian~\cite{tsitsulin18kdd}.}}
% Mind that the graph's Laplacian is just an operator, not a representation -
% END of Page 2, 01dec2019

%\input{tables/tbl-notation}
%!TEX root=../main.tex
\section{Solution}\label{sec:solution}
% Story 1 
% - Let us start our discussion with assuming that we are given a bijective mapping between the nodes of the graph
% - If we had such function we could apply the indicator function on nodes of graph G_2 to obtain nodes on the graph G_2
% - This reasoning work on any real valued function f,g on nodes in G_1, G_2. b7y8
% - That means that if we know a set of functions which corresponds one to another we could recover the behavior of the indicator functions on the nodes.
% - However, there are two problems: not all functions can act as good corresponding functions, and not all represent the behaviour of the indicator function. 
% - Can we say here that Heat is a good choice as the indicator function can be immediately represented as H_{0,i}? 
% - So if we are able to align functions on G_1 and G_2 we will be able to align nodes in the two graphs by means of representations over the nodes (Good!)
% Now how do we find C? 
% What happens if graphs are not close to be isomorphic? Base alignment

\pkw{Here, we choose an orthonormal basis and a function, which are, in our judgement, appropriate for node alignment purposes, and define the complete pipeline of our solution.}
% pk clarification of choice 20oct2020

\subsection{Choice of basis: Normalized Laplacian}

% WHAT? paragraph
\pkw{As a basis for representing functions as linear combinations of base functions, we use} the eigenvectors of the graph's \emph{normalized Laplacian}, i.e., the matrix $\lapl=\eye-D^{-\frac{1}{2}}AD^{-\frac{1}{2}}$, where $D$ a diagonal degree matrix of node degrees $D_{ii} = \sum_{j=1}^n A_{ij}$ and $A$ is the graph adjacency matrix; its eigendecomposition is  $\lapl=\Phi\Lambda\Phi^\top$, where $\Lambda$ is a diagonal matrix of eigenvalues, $\{\lambda_1, \ldots, \lambda_n\}$, i.e., the graph's \emph{spectrum}, which encodes information about communities, degree distribution, and diameter, and~$\Phi$ is a matrix of eigenvectors, $\Phi_\lapl = [\phi_1 \phi_2 \ldots \phi_n]$. The eigenvectors form an orthogonal basis, which we use a standard basis. \pk{We use~$\phi$ ($\psi$) to indicate the eigenvectors of the Laplacian of graph~$G_1$ ($G_2$).}
% condensed 20jan2021

% WHY? paragraph
\pkw{We consider this basis to be suitable, since the eigenvectors of the normalized Laplacian converge to the eigenfunctions of the Laplace-Berltrami operator~\cite{belkin06}, which measures the smoothness of continuous surfaces.}

\subsection{Choice of function: Heat Kernel}

\pkw{The choice of functions $f_i: \V_1 \rightarrow \reals$, $g_i: \V_2 \rightarrow \reals$, is critical for our method. A poor choice would be detrimental. A good choice should have the following properties:}
% condensed 20jan2021

% DM: Trying to address Rev. 2 concerns
\pkw{\spara{Expressiveness.} The function should express the graph's structure. For instance, a constant function returning the same value for all nodes would not yield a meaningful alignment.}

\pkw{\spara{Permutation-invariance.} The function should not depend on the node index $i$; the indicator function \pki{lacks} this property.}
%\dmt{\spara{Compactness.} A function is compact if it can be easily approximated by a small $k$ number of basis functions. To be compact $k\ll n$ should be independent from the number of nodes.}

% why did we try to tone down the multiscale property?
\pkw{\spara{Multiscale robustness.} \pki{The function should robustly capture both local and global structures (e.g., edges and communities), insensitively to small perturbations.}}

\smallskip
\pk{\pkw{A function fulfilling these requirements is the time-parameterized} \emph{heat kernel}~\cite{tsitsulin18kdd}:
% condensed 20jan2021

\vspace{-3mm}
\begin{equation}\label{eq:heat-kernel}
\textstyle H_t = \Phi e^{-t\Lambda}\Phi^\top = \sum_{j=1}^n e^{-t\lambda_j}\phi_j\phi_j^\top
\end{equation}
\vspace{-4mm}

\noindent where $H_{t[ij]}$ measures the flow of heat from node~$i$ to node~$j$ at time~$t$, as it diffuses from \pkw{each} node's neighborhood to the whole graph. We build our model functions over a sequence of time steps~$t$ using the \emph{diagonal} of the heat kernel, which measures the heat flowing back to each node at time~$t$. 
 
% values become *relatively* smaller or larger, wrt each other
\pkw{The heat kernel %enjoys the aforementioned properties: it
expresses \pki{multiscale} graph structure in a permutation-invariant manner and is robust to small changes. %, as the value in the diagonal can be thought as the likelihood a node is reached by a diffusion process within time~$t$.
In the beginning of the diffusion, Equation~\eqref{eq:heat-kernel} emphasises large~$\lambda$, which correspond to \emph{local} edge and ego-net properties. As time progresses, smaller eigenvalues get emphasized, reflecting \emph{global} graph properties, such as communities.}}
%We recall that $\phi_2$ is the Fiedler vector used in community detection.

% This seems out of place. Why do we need to call them corresponding functions and confuse the reader with the use of correspondence matrix
\smallskip
\pk{\pkw{We build our \emph{corresponding functions} $f_i$, $g_i$, from the heat kernel at different time steps $t$,} as linear combinations of the graph's Laplacian orthogonal eigenvectors.
%, i.e., $f_i = \sum_{j=1}^n a_{ij}\phi_j$. % confusing and contradictory
Let~$F \in \reals^{n\times q}$, $F = [f_1, \ldots, f_q]$ be the matrix containing the diagonals of the heat kernel of~$G_1$, $H_t^{G_1}$, over~$q$ time\footnote{In our experiments we select $q=100$ values evenly spaced on the linear scale in the range [0.1, 50].} steps, ${f_i= \sum_{j=1}^n e^{-{t_i}\lambda_j}\phi_j \odot \phi_j}$, where~$\odot$ denotes the element-wise vector product. Likewise, the matrix $G \in \reals^{n\times q}, G = [g_1, \ldots, g_q]$ contains the diagonals of~$H_t^{G_2}$, the heat kernel of~$G_2$. While the~$q$ \emph{columns} of~$F$ and~$G$ contain the same time-dependent heat-kernel-diagonal functions on the nodes of two graphs, their~$n$ \emph{rows} (i.e., nodes) are not aligned. We need to obtain such a node alignment.}
% pk edited and simplified, aligned, 20oct2020

\subsection{Mapping matrix}\label{sec:matrix_C}

\pk{We approximate each function $f_i$ using only the first $k$ eigenvectors, as done, by analogy, 
%with eigenvectors of the Laplace-Beltrami operator 
on shapes analysis~\cite{belkin06},
%for Fourier series expansion in the manner of a low-pass filter~\cmt{REF}. % is this needed here?
and thereby calculate the corresponding function matrices $F$ and $G$.}
\dm{$F$ and $G$ can be thought as coefficient matrices \pk{used to produce} linear combinations, $F^\top\Phi$ and $G^\top\Psi$, of the Laplacian eigenvectors of $G_1$ and $G_2$, respectively. With a slight abuse of notation, we denote with $\Phi$ and $\Psi$ the first $k$ eigenvectors, \pk{hence $F^\top\Phi$ and $G^\top\Psi$ are in $\reals^{q\times k}$.} In the projection of the functions on the first $k$ eigenvectors, we would like the corresponding functions to be equal up to a coefficient matrix $C \!\in\! \reals^{k \times k}$. In the case of isomorphic graphs, it holds that $F^\top\Phi=G^\top\Psi C$%, where~\pkw{$C$ is a diagonal mapping matrix}, hence:

\vspace{-2mm}
\begin{equation}
\resizebox{.5\columnwidth}{!}{$
    \begin{bmatrix}
    \diag(g_1^\top\Psi)\\
    \vdots\\
    \diag(g_q^\top\Psi)
\end{bmatrix}
    \begin{bmatrix}
    c_{11}\\
    \vdots\\
    c_{kk}
\end{bmatrix}
=    \begin{bmatrix}
    \Phi^\top f_1\\
    \vdots\\
    \Phi^\top f_q
\end{bmatrix}
$}
\label{eq:corrmat}
\end{equation}
\vspace{-1mm}

% therefore, every function $f_i$ is composed as $\Phi f_i = C $
% From these \emph{coefficient matrices} \dm{$A,B \in \reals^{k\times q}$}, such that  $f_i = \sum_{j=1}^k A_{ij}\phi_j$ and $g_i = \sum_{j=1}^k B_{ij}\jh{\psi}_j$ can be calculated, from which we aim to derive the correspondence matrix $C \in \reals^{k\times k}$ as $B=CA$. 
%Instead of solving for $C$ here, % absurd statement!
\pk{Matrix~$C$ \pkw{is diagonal in the case of isomorphic graphs and deviates from a diagonal form as graphs diverge from isomorphism; for simplicity, we assume} a diagonal~$C$, and obtain the diagonal \pkw{entries} that minimize} the~$L_2$-norm difference $\|\cdot\|^2_2$ between the left and rights side of Eq.~\eqref{eq:corrmat} \pk{using the ordinary least squares method, as in~\cite{kovnatsky13coupled}.} \pkw{In Section~\ref{ssec:basealign} we delve into the case of non-isomorphic graphs.}} % condensed 20jan2021

\subsection{Node-to-node correspondence}\label{ssec:matching}

\dm{
% \begin{itemize}
%     \item Look for functions with same coefficients
%     \item Easy function: delta functions $\rightarrow$ what are their coefficients
%     \item heat kernel row at time $t=0$
%     \item imagine as matching of node descriptors
% \end{itemize}

% redundant statement, which is repeated later.
%In order to compute the node-to-node correspondence for graph alignment, we exploit the fact that in the ideal case \emph{corresponding functions have the same coefficients.}
We consider the delta function $\delta_i(\cdot)$ %introduced in Section~\ref{sec:maps} % broken link
as corresponding function; these functions yield an $n\times n$ identity matrix. We express such a function as a vector of coefficients, since the vector of $\delta_i$ is the $i$th row of the heat kernel at $t=0$:

\vspace{-3mm}
\[\textstyle\delta_i=H_{i,t=0}^{G_1}=\sum_{j=1}^n \phi_{ij}\phi_j\]
\vspace{-4mm}

The computation for delta functions on $G_2$ follows equivalently using
$\Psi$ in place of $\Phi$.
%$\hat{\Psi}$ instead of $\bar{\Phi}$. % not introduced!
We may match the coefficient vectors of these corresponding indicator functions, as, \pk{ideally,} for two matching nodes $v_i\in V_1$ and $v'_j\in V_2$, the coefficients of $\delta_i$ and $\delta_j$ for $\Phi$ and $\Psi$
%$\bar{\Phi}$ and $\hat{\Psi}$ bar/hat not introduced!
should be \pk{identical}. \pk{In particular,} the coefficients expressing $\delta_i$ as a linear combination of the first $k$ eigenvectors are $\phi_{i1}, \ldots, \phi_{ik}$. We set
$\Phi^\top$ and $C\Psi^\top$ in $\reals^{k\times n}$
%$\bar{\Phi}^\top$ and $C\hat{\Psi}^\top$ bar/hat notations never introduced!
as coefficient matrices of the delta functions, aligned by~$C$. Rows correspond to the first $k$ Laplacian eigenvectors, while \pk{columns stand for graph nodes, rather than for time steps of heat diffusion}. We need to match coefficient vectors, i.e., columns of $\Phi^\top$ and $C\Psi^\top$,
%$\bar{\Phi}^\top$ and $C\hat{\Psi}^\top$, % undefined!
to each other. This problem amounts to a \emph{linear assignment problem}; we apply an \pk{off-the-shelf algorithm therefore}, such as \textbf{nearest neighbor search} or \textbf{Jonker-Volgenant (JV)}~\cite{jonker1987shortest}, to obtain a one-to-one matching between the columns of $\Phi^\top$ and $C\Psi^\top$,
%$\bar{\Phi}^\top$ and $C\hat{\Psi}^\top$, % not defined!
and hence an alignment of nodes.}
% then we use a \emph{linear assignment algorithm} such as \textbf{nearest neighbor search} or \textbf{Jonker-Volgenant (JV)~\cite{jonker1987shortest}} to obtain an one-to-one matching between the columns of $C F^\delta$ and those of $G^\delta$ \cmt{what objective function is minimized here?}, and hence an alignment of nodes in $G_1$ to nodes in $G_2$.}
\pki{\thiswork is flexible in that we may choose any matching method.}
% as we may employ a different matching algorithm and a different transformation on the coefficient matrix, while preserving the overall framework.} 
%Section~\ref{ssec:variants} compare variants using nearest neighbor or JV. % broken link
%SortGreedy~\cite{doka2015k}

\subsection{Base Alignment}\label{ssec:basealign}

%!TEX root=../ijcai21.tex

% \begin{figure*}[htpb]
%   \setbox9=\hbox{\includegraphics[width=.3\linewidth]{example-image-1x1}}% Capture tallest image in box 9
%   \subcaptionbox{Curl-free potential singularities}
%     {\raisebox{\dimexpr\ht9-\height}{\includegraphics[width=.3\linewidth]{example-image-a}}} \hfill
%   \subcaptionbox{Divergence-free potential singularities}
%     {\raisebox{\dimexpr\ht9-\height}{\includegraphics[width=.3\linewidth]{example-image-b}}} \hfill
%   \subcaptionbox{Vector field potentials singularities}{\includegraphics[width=.3\linewidth]
%     {example-image-1x1}}
% \end{figure*}

\begin{figure}[!t]
\centering
\subcaptionbox{Two graphs spectra}[.35\linewidth]{
\centering
\resizebox{.35\linewidth}{!}{%!TEX root=../ijcai21.tex

\begin{tikzpicture}
	\begin{axis}[
		ylabel={\Large Eigenvalue $\lambda_i$},
		xlabel={\Large $i$},
		xmin=0,
		xmax=33,
%		xmode=log,
		ymin=0,
		ymax=2,
%		legend cell align=left,
%		legend pos = north east,
%		extra x ticks={50000},
		% width=.3\linewidth,
		height=4.5cm,
		every tick label/.append style={font=\large}
	]
	\addplot[thick,color=cycle3,mark=*,mark size=1pt] table[x=idx,y=karate] {data/karate_spectrum.dat};
	\addplot[thick,color=cycle2,mark=*,mark size=1pt] table[x=idx,y=new] {data/karate_spectrum.dat};
	\node at (10,1.5) {\resizebox{1.75cm}{!}{\begin{tikzpicture}
\foreach \nodename/\x/\y/\nc in {
0/8.212188782135431/-0.352861046224822, 1/6.55641100473977/-0.00389157237775555, 2/-0.6335245434172445/-0.0030022048501545584, 3/3.810046985633419/-0.42440486692447055, 4/12.265822255747468/-0.9125437625311121, 5/18.07949532854742/-0.7922116673728108, 6/16.15594342882408/-0.9524317351396732, 7/5.246835676220883/-0.259183814470877, 8/-0.5581927708086516/0.18861369389277002, 9/-0.9718606916302384/0.6094182660158936, 10/14.654205501087578/-0.7033538780130998, 11/18.40124075476423/-0.3188429363775077, 12/5.91829585497284/-0.8366708290161182, 13/1.6422185729523857/-0.09058293537259482, 14/-12.492778834813304/0.7921961912453461, 15/-10.798663215811121/0.94003704411126, 16/22.499999999999996/-1.1656824194978666, 17/13.270697902054597/0.07332139518510909, 18/-8.0494814015231/0.9504031469012817, 19/3.7368254216944683/0.20763961761204003, 20/-15.358082338904119/0.42343546510338537, 21/13.630002073637122/-0.14246408764063334, 22/-5.226217569736741/0.9215458288389458, 23/-13.584126306023668/0.1749992453184146, 24/-12.51733632823277/-0.48647141297047347, 25/-14.903076744306983/-0.2829895382405123, 26/-16.846586581121535/0.7640177940970034, 27/-9.956649325731645/-0.08371887114770163, 28/-6.679840689864612/0.05591492320529544, 29/-14.083424546592562/0.556423136204105, 30/-0.48246376292148746/0.41672543960441866, 31/-5.860857601512076/-0.15842607808486453, 32/-8.023272132708334/0.4822635400800517, 33/-7.053794157351485/0.4127789288377293
}
{
\node (\nodename) at (\x,\y) [shape=circle,inner sep=3pt,draw,thick,fill=cycle3] {}; % cycle2 or cycle5
}
\begin{pgfonlayer}{background}
\path
\foreach \startnode/\endnode in {
2/0, 2/1, 2/3, 2/7, 2/8, 2/9, 2/13, 2/27, 2/28, 2/32, 0/3, 0/4, 0/5, 0/6, 0/7, 0/10, 0/11, 0/12, 0/17, 0/19, 0/21, 0/31, 3/1, 3/7, 4/6, 4/10, 5/6, 5/10, 5/16, 6/16, 7/1, 19/1, 19/33, 21/1, 31/24, 31/28, 31/33, 1/13, 1/30, 13/33, 30/8, 30/32, 30/33, 8/32, 8/33, 9/33, 27/23, 27/33, 28/33, 32/14, 32/15, 32/20, 32/22, 32/23, 32/29, 33/15, 33/18, 33/20, 33/22, 33/26, 33/29, 25/23, 25/24, 23/29, 29/26}
{
(\startnode) edge[-,thick,draw opacity=0.75] node {} (\endnode)
};
\path
\foreach \startnode/\endnode in {
0/1, 0/8, 0/13, 1/17, 3/12, 3/13, 14/33, 18/32, 23/33, 24/27, 25/31, 31/32, 32/33}
{
(\startnode) edge[-,ultra thick, draw=cycle1] node {} (\endnode)
};
\end{pgfonlayer}
\end{tikzpicture}}};
	\node at (25,0.5) {\resizebox{1.75cm}{!}{\begin{tikzpicture}
\foreach \nodename/\x/\y/\nc in {
0/8.212188782135431/-0.352861046224822, 1/6.55641100473977/-0.00389157237775555, 2/-0.6335245434172445/-0.0030022048501545584, 3/3.810046985633419/-0.42440486692447055, 4/12.265822255747468/-0.9125437625311121, 5/18.07949532854742/-0.7922116673728108, 6/16.15594342882408/-0.9524317351396732, 7/5.246835676220883/-0.259183814470877, 8/-0.5581927708086516/0.18861369389277002, 9/-0.9718606916302384/0.6094182660158936, 10/14.654205501087578/-0.7033538780130998, 11/18.40124075476423/-0.3188429363775077, 12/5.91829585497284/-0.8366708290161182, 13/1.6422185729523857/-0.09058293537259482, 14/-12.492778834813304/0.7921961912453461, 15/-10.798663215811121/0.94003704411126, 16/22.499999999999996/-1.1656824194978666, 17/13.270697902054597/0.07332139518510909, 18/-8.0494814015231/0.9504031469012817, 19/3.7368254216944683/0.20763961761204003, 20/-15.358082338904119/0.42343546510338537, 21/13.630002073637122/-0.14246408764063334, 22/-5.226217569736741/0.9215458288389458, 23/-13.584126306023668/0.1749992453184146, 24/-12.51733632823277/-0.48647141297047347, 25/-14.903076744306983/-0.2829895382405123, 26/-16.846586581121535/0.7640177940970034, 27/-9.956649325731645/-0.08371887114770163, 28/-6.679840689864612/0.05591492320529544, 29/-14.083424546592562/0.556423136204105, 30/-0.48246376292148746/0.41672543960441866, 31/-5.860857601512076/-0.15842607808486453, 32/-8.023272132708334/0.4822635400800517, 33/-7.053794157351485/0.4127789288377293
}
{
\node (\nodename) at (\x,\y) [shape=circle,inner sep=3pt,draw,thick,fill=cycle2] {}; % cycle2 or cycle5
}
\begin{pgfonlayer}{background}
\path
\foreach \startnode/\endnode in {
2/0, 2/1, 2/3, 2/7, 2/8, 2/9, 2/13, 2/27, 2/28, 2/32, 0/3, 0/4, 0/5, 0/6, 0/7, 0/10, 0/11, 0/12, 0/17, 0/19, 0/21, 0/31, 3/1, 3/7, 4/6, 4/10, 5/6, 5/10, 5/16, 6/16, 7/1, 19/1, 19/33, 21/1, 31/24, 31/28, 31/33, 1/13, 1/30, 13/33, 30/8, 30/32, 30/33, 8/32, 8/33, 9/33, 27/23, 27/33, 28/33, 32/14, 32/15, 32/20, 32/22, 32/23, 32/29, 33/15, 33/18, 33/20, 33/22, 33/26, 33/29, 25/23, 25/24, 23/29, 29/26}
{
(\startnode) edge[-,thick,draw opacity=0.75] node {} (\endnode)
};
\end{pgfonlayer}
\end{tikzpicture}}};
	\end{axis}
\end{tikzpicture}}
}
\hfill
\subcaptionbox{Their first 3 eigenvectors}[.63\linewidth]{
\centering
\resizebox{.2\linewidth}{!}{\begin{tikzpicture}
\foreach \nodename/\x/\y/\cp/\cn/\cw in {
0/1.46/-0.71/0.0000/0.8557/0.1443, 1/1.17/-0.01/0.0000/0.3274/0.6726, 2/-0.11/-0.01/0.0333/0.0000/0.9667, 3/0.68/-0.85/0.0000/0.3324/0.6676, 4/2.18/-1.83/0.0000/0.7713/0.2287, 5/3.21/-1.58/0.0000/1.0000/0.0000, 6/2.87/-1.90/0.0000/1.0000/0.0000, 7/0.93/-0.52/0.0000/0.2596/0.7404, 8/-0.10/0.38/0.1958/0.0000/0.8042, 9/-0.17/1.22/0.2062/0.0000/0.7938, 10/2.61/-1.41/0.0000/0.7713/0.2287, 11/3.27/-0.64/0.0000/0.2465/0.7535, 12/1.05/-1.67/0.0000/0.2849/0.7151, 13/0.29/-0.18/0.0000/0.1349/0.8651, 14/-2.22/1.58/0.4170/0.0000/0.5830, 15/-1.92/1.88/0.4170/0.0000/0.5830, 16/4.00/-2.33/0.0000/0.8149/0.1851, 17/2.36/0.15/0.0000/0.2633/0.7367, 18/-1.43/1.90/0.4170/0.0000/0.5830, 19/0.66/0.42/0.0000/0.0893/0.9107, 20/-2.73/0.85/0.4170/0.0000/0.5830, 21/2.42/-0.28/0.0000/0.2633/0.7367, 22/-0.93/1.84/0.4170/0.0000/0.5830, 23/-2.41/0.35/0.7266/0.0000/0.2734, 24/-2.23/-0.97/0.5020/0.0000/0.4980, 25/-2.65/-0.57/0.5380/0.0000/0.4620, 26/-2.99/1.53/0.4725/0.0000/0.5275, 27/-1.77/-0.17/0.5001/0.0000/0.4999, 28/-1.19/0.11/0.2974/0.0000/0.7026, 29/-2.50/1.11/0.6747/0.0000/0.3253, 30/-0.09/0.83/0.2646/0.0000/0.7354, 31/-1.04/-0.32/0.4748/0.0000/0.5252, 32/-1.43/0.96/0.9327/0.0000/0.0673, 33/-1.25/0.83/1.0000/0.0000/0.0000
}
{
\node (\nodename) at (\x,\y) [shape=circle,inner sep=3pt,draw,thick,fill={rgb:cycle1,\cp;white,\cw;cycle2,\cn}] {}; % cycle2 or cycle5
}
\begin{pgfonlayer}{background}
\path
\foreach \startnode/\endnode in {
2/0, 2/1, 2/3, 2/7, 2/8, 2/9, 2/13, 2/27, 2/28, 2/32, 0/3, 0/4, 0/5, 0/6, 0/7, 0/10, 0/11, 0/12, 0/17, 0/19, 0/21, 0/31, 3/1, 3/7, 4/6, 4/10, 5/6, 5/10, 5/16, 6/16, 7/1, 19/1, 19/33, 21/1, 31/24, 31/28, 31/33, 1/13, 1/30, 13/33, 30/8, 30/32, 30/33, 8/32, 8/33, 9/33, 27/23, 27/33, 28/33, 32/14, 32/15, 32/20, 32/22, 32/23, 32/29, 33/15, 33/18, 33/20, 33/22, 33/26, 33/29, 25/23, 25/24, 23/29, 29/26}
{
(\startnode) edge[-,thick,draw opacity=0.75] node {} (\endnode)
};
\path
\foreach \startnode/\endnode in {
0/1, 0/8, 0/13, 1/17, 3/12, 3/13, 14/33, 18/32, 23/33, 24/27, 25/31, 31/32, 32/33}
{
(\startnode) edge[-,ultra thick, draw=cycle1] node {} (\endnode)
};
\node (name) at (0, -2.5) [scale=2.5] {\Large $\phi_1$};
\end{pgfonlayer}
\end{tikzpicture}}\hfill
\resizebox{.2\linewidth}{!}{\begin{tikzpicture}
\foreach \nodename/\x/\y/\cp/\cn/\cw in {
0/1.46/-0.71/0.4126/0.0000/0.5874, 1/1.17/-0.01/1.0000/0.0000/0.0000, 2/-0.11/-0.01/0.6034/0.0000/0.3966, 3/0.68/-0.85/0.8962/0.0000/0.1038, 4/2.18/-1.83/0.0000/0.5868/0.4132, 5/3.21/-1.58/0.0000/1.0000/0.0000, 6/2.87/-1.90/0.0000/1.0000/0.0000, 7/0.93/-0.52/0.6965/0.0000/0.3035, 8/-0.10/0.38/0.1386/0.0000/0.8614, 9/-0.17/1.22/0.1259/0.0000/0.8741, 10/2.61/-1.41/0.0000/0.5868/0.4132, 11/3.27/-0.64/0.1447/0.0000/0.8553, 12/1.05/-1.67/0.4652/0.0000/0.5348, 13/0.29/-0.18/0.5829/0.0000/0.4171, 14/-2.22/1.58/0.0000/0.1721/0.8279, 15/-1.92/1.88/0.0000/0.1721/0.8279, 16/4.00/-2.33/0.0000/0.9918/0.0082, 17/2.36/0.15/0.4329/0.0000/0.5671, 18/-1.43/1.90/0.0000/0.1721/0.8279, 19/0.66/0.42/0.3017/0.0000/0.6983, 20/-2.73/0.85/0.0000/0.1721/0.8279, 21/2.42/-0.28/0.4329/0.0000/0.5671, 22/-0.93/1.84/0.0000/0.1721/0.8279, 23/-2.41/0.35/0.0000/0.4968/0.5032, 24/-2.23/-0.97/0.0000/0.4490/0.5510, 25/-2.65/-0.57/0.0000/0.5056/0.4944, 26/-2.99/1.53/0.0000/0.2756/0.7244, 27/-1.77/-0.17/0.0000/0.2391/0.7609, 28/-1.19/0.11/0.0000/0.0020/0.9980, 29/-2.50/1.11/0.0000/0.4141/0.5859, 30/-0.09/0.83/0.1673/0.0000/0.8327, 31/-1.04/-0.32/0.0000/0.3502/0.6498, 32/-1.43/0.96/0.0000/0.3558/0.6442, 33/-1.25/0.83/0.0000/0.2918/0.7082
}
{
\node (\nodename) at (\x,\y) [shape=circle,inner sep=3pt,draw,thick,fill={rgb:cycle1,\cp;white,\cw;cycle2,\cn}] {}; % cycle2 or cycle5
}
\begin{pgfonlayer}{background}
\path
\foreach \startnode/\endnode in {
2/0, 2/1, 2/3, 2/7, 2/8, 2/9, 2/13, 2/27, 2/28, 2/32, 0/3, 0/4, 0/5, 0/6, 0/7, 0/10, 0/11, 0/12, 0/17, 0/19, 0/21, 0/31, 3/1, 3/7, 4/6, 4/10, 5/6, 5/10, 5/16, 6/16, 7/1, 19/1, 19/33, 21/1, 31/24, 31/28, 31/33, 1/13, 1/30, 13/33, 30/8, 30/32, 30/33, 8/32, 8/33, 9/33, 27/23, 27/33, 28/33, 32/14, 32/15, 32/20, 32/22, 32/23, 32/29, 33/15, 33/18, 33/20, 33/22, 33/26, 33/29, 25/23, 25/24, 23/29, 29/26}
{
(\startnode) edge[-,thick,draw opacity=0.75] node {} (\endnode)
};
\path
\foreach \startnode/\endnode in {
0/1, 0/8, 0/13, 1/17, 3/12, 3/13, 14/33, 18/32, 23/33, 24/27, 25/31, 31/32, 32/33}
{
(\startnode) edge[-,ultra thick, draw=cycle1] node {} (\endnode)
};
\node (name) at (0, -2.5) [scale=2.5] {\Large $\phi_2$};
\end{pgfonlayer}
\end{tikzpicture}}\hfill
\resizebox{.2\linewidth}{!}{\begin{tikzpicture}
\foreach \nodename/\x/\y/\cp/\cn/\cw in {
0/1.46/-0.71/0.1222/0.0000/0.8778, 1/1.17/-0.01/0.0232/0.0000/0.9768, 2/-0.11/-0.01/0.0937/0.0000/0.9063, 3/0.68/-0.85/0.1247/0.0000/0.8753, 4/2.18/-1.83/0.0000/0.0722/0.9278, 5/3.21/-1.58/0.0000/0.1872/0.8128, 6/2.87/-1.90/0.0000/0.1872/0.8128, 7/0.93/-0.52/0.0969/0.0000/0.9031, 8/-0.10/0.38/0.0000/0.4805/0.5195, 9/-0.17/1.22/0.0000/0.1686/0.8314, 10/2.61/-1.41/0.0000/0.0722/0.9278, 11/3.27/-0.64/0.0499/0.0000/0.9501, 12/1.05/-1.67/0.0940/0.0000/0.9060, 13/0.29/-0.18/0.0095/0.0000/0.9905, 14/-2.22/1.58/0.0000/0.5674/0.4326, 15/-1.92/1.88/0.0000/0.5674/0.4326, 16/4.00/-2.33/0.0000/0.2160/0.7840, 17/2.36/0.15/0.0442/0.0000/0.9558, 18/-1.43/1.90/0.0000/0.5674/0.4326, 19/0.66/0.42/0.0000/0.1220/0.8780, 20/-2.73/0.85/0.0000/0.5674/0.4326, 21/2.42/-0.28/0.0442/0.0000/0.9558, 22/-0.93/1.84/0.0000/0.5674/0.4326, 23/-2.41/0.35/0.2977/0.0000/0.7023, 24/-2.23/-0.97/1.0000/0.0000/0.0000, 25/-2.65/-0.57/0.9233/0.0000/0.0767, 26/-2.99/1.53/0.0000/0.5178/0.4822, 27/-1.77/-0.17/0.5177/0.0000/0.4823, 28/-1.19/0.11/0.1821/0.0000/0.8179, 29/-2.50/1.11/0.0000/0.4914/0.5086, 30/-0.09/0.83/0.0000/0.5645/0.4355, 31/-1.04/-0.32/0.6597/0.0000/0.3403, 32/-1.43/0.96/0.0000/1.0000/0.0000, 33/-1.25/0.83/0.0000/0.8368/0.1632
}
{
\node (\nodename) at (\x,\y) [shape=circle,inner sep=3pt,draw,thick,fill={rgb:cycle1,\cp;white,\cw;cycle2,\cn}] {}; % cycle2 or cycle5
}
\begin{pgfonlayer}{background}
\path
\foreach \startnode/\endnode in {
2/0, 2/1, 2/3, 2/7, 2/8, 2/9, 2/13, 2/27, 2/28, 2/32, 0/3, 0/4, 0/5, 0/6, 0/7, 0/10, 0/11, 0/12, 0/17, 0/19, 0/21, 0/31, 3/1, 3/7, 4/6, 4/10, 5/6, 5/10, 5/16, 6/16, 7/1, 19/1, 19/33, 21/1, 31/24, 31/28, 31/33, 1/13, 1/30, 13/33, 30/8, 30/32, 30/33, 8/32, 8/33, 9/33, 27/23, 27/33, 28/33, 32/14, 32/15, 32/20, 32/22, 32/23, 32/29, 33/15, 33/18, 33/20, 33/22, 33/26, 33/29, 25/23, 25/24, 23/29, 29/26}
{
(\startnode) edge[-,thick,draw opacity=0.75] node {} (\endnode)
};
\path
\foreach \startnode/\endnode in {
0/1, 0/8, 0/13, 1/17, 3/12, 3/13, 14/33, 18/32, 23/33, 24/27, 25/31, 31/32, 32/33}
{
(\startnode) edge[-,ultra thick, draw=cycle1] node {} (\endnode)
};
\node (name) at (0, -2.5) [scale=2.5] {\Large $\phi_3$};
\end{pgfonlayer}
\end{tikzpicture}}
\resizebox{.2\linewidth}{!}{\begin{tikzpicture}
\foreach \nodename/\x/\y/\cp/\cn/\cw in {
0/1.46/-0.71/0.0000/0.9508/0.0492, 1/1.17/-0.01/0.0000/0.0239/0.9761, 2/-0.11/-0.01/0.2412/0.0000/0.7588, 3/0.68/-0.85/0.0000/0.1712/0.8288, 4/2.18/-1.83/0.0000/0.7909/0.2091, 5/3.21/-1.58/0.0000/1.0000/0.0000, 6/2.87/-1.90/0.0000/1.0000/0.0000, 7/0.93/-0.52/0.0000/0.1712/0.8288, 8/-0.10/0.38/0.5082/0.0000/0.4918, 9/-0.17/1.22/0.2680/0.0000/0.7320, 10/2.61/-1.41/0.0000/0.7909/0.2091, 11/3.27/-0.64/0.0000/0.2960/0.7040, 12/1.05/-1.67/0.0000/0.2960/0.7040, 13/0.29/-0.18/0.2104/0.0000/0.7896, 14/-2.22/1.58/0.3742/0.0000/0.6258, 15/-1.92/1.88/0.4720/0.0000/0.5280, 16/4.00/-2.33/0.0000/0.7938/0.2062, 17/2.36/0.15/0.0000/0.2960/0.7040, 18/-1.43/1.90/0.2933/0.0000/0.7067, 19/0.66/0.42/0.0000/0.0585/0.9415, 20/-2.73/0.85/0.4720/0.0000/0.5280, 21/2.42/-0.28/0.0000/0.2165/0.7835, 22/-0.93/1.84/0.4720/0.0000/0.5280, 23/-2.41/0.35/0.7277/0.0000/0.2723, 24/-2.23/-0.97/0.3234/0.0000/0.6766, 25/-2.65/-0.57/0.4703/0.0000/0.5297, 26/-2.99/1.53/0.4997/0.0000/0.5003, 27/-1.77/-0.17/0.4546/0.0000/0.5454, 28/-1.19/0.11/0.2673/0.0000/0.7327, 29/-2.50/1.11/0.7363/0.0000/0.2637, 30/-0.09/0.83/0.4692/0.0000/0.5308, 31/-1.04/-0.32/0.1497/0.0000/0.8503, 32/-1.43/0.96/1.0000/0.0000/0.0000, 33/-1.25/0.83/0.9778/0.0000/0.0222
}
{
\node (\nodename) at (\x,\y) [shape=circle,inner sep=3pt,draw,thick,fill={rgb:cycle1,\cp;white,\cw;cycle2,\cn}] {}; % cycle2 or cycle5
}
\begin{pgfonlayer}{background}
\path
\foreach \startnode/\endnode in {
2/0, 2/1, 2/3, 2/7, 2/8, 2/9, 2/13, 2/27, 2/28, 2/32, 0/3, 0/4, 0/5, 0/6, 0/7, 0/10, 0/11, 0/12, 0/17, 0/19, 0/21, 0/31, 3/1, 3/7, 4/6, 4/10, 5/6, 5/10, 5/16, 6/16, 7/1, 19/1, 19/33, 21/1, 31/24, 31/28, 31/33, 1/13, 1/30, 13/33, 30/8, 30/32, 30/33, 8/32, 8/33, 9/33, 27/23, 27/33, 28/33, 32/14, 32/15, 32/20, 32/22, 32/23, 32/29, 33/15, 33/18, 33/20, 33/22, 33/26, 33/29, 25/23, 25/24, 23/29, 29/26}
{
(\startnode) edge[-,thick,draw opacity=0.75] node {} (\endnode)
};
\node (name) at (0, -2.5) [scale=2.5] {\Large $\psi_1$};
\end{pgfonlayer}
\end{tikzpicture}}\hfill
\resizebox{.2\linewidth}{!}{\begin{tikzpicture}
\foreach \nodename/\x/\y/\cp/\cn/\cw in {
0/1.46/-0.71/0.2232/0.0000/0.7768, 1/1.17/-0.01/1.0000/0.0000/0.0000, 2/-0.11/-0.01/0.6795/0.0000/0.3205, 3/0.68/-0.85/0.7103/0.0000/0.2897, 4/2.18/-1.83/0.0000/0.2513/0.7487, 5/3.21/-1.58/0.0000/0.4306/0.5694, 6/2.87/-1.90/0.0000/0.4306/0.5694, 7/0.93/-0.52/0.7103/0.0000/0.2897, 8/-0.10/0.38/0.2261/0.0000/0.7739, 9/-0.17/1.22/0.2208/0.0000/0.7792, 10/2.61/-1.41/0.0000/0.2513/0.7487, 11/3.27/-0.64/0.0871/0.0000/0.9129, 12/1.05/-1.67/0.0871/0.0000/0.9129, 13/0.29/-0.18/0.4872/0.0000/0.5128, 14/-2.22/1.58/0.0000/0.0649/0.9351, 15/-1.92/1.88/0.0000/0.0400/0.9600, 16/4.00/-2.33/0.0000/0.4283/0.5717, 17/2.36/0.15/0.0871/0.0000/0.9129, 18/-1.43/1.90/0.0100/0.0000/0.9900, 19/0.66/0.42/0.3630/0.0000/0.6370, 20/-2.73/0.85/0.0000/0.0400/0.9600, 21/2.42/-0.28/0.4375/0.0000/0.5625, 22/-0.93/1.84/0.0000/0.0400/0.9600, 23/-2.41/0.35/0.0000/0.7176/0.2824, 24/-2.23/-0.97/0.0000/0.9144/0.0856, 25/-2.65/-0.57/0.0000/1.0000/0.0000, 26/-2.99/1.53/0.0000/0.1777/0.8223, 27/-1.77/-0.17/0.0000/0.1429/0.8571, 28/-1.19/0.11/0.0000/0.0239/0.9761, 29/-2.50/1.11/0.0000/0.3690/0.6310, 30/-0.09/0.83/0.3109/0.0000/0.6891, 31/-1.04/-0.32/0.0000/0.4245/0.5755, 32/-1.43/0.96/0.0000/0.1384/0.8616, 33/-1.25/0.83/0.0267/0.0000/0.9733
}
{
\node (\nodename) at (\x,\y) [shape=circle,inner sep=3pt,draw,thick,fill={rgb:cycle1,\cp;white,\cw;cycle2,\cn}] {}; % cycle2 or cycle5
}
\begin{pgfonlayer}{background}
\path
\foreach \startnode/\endnode in {
2/0, 2/1, 2/3, 2/7, 2/8, 2/9, 2/13, 2/27, 2/28, 2/32, 0/3, 0/4, 0/5, 0/6, 0/7, 0/10, 0/11, 0/12, 0/17, 0/19, 0/21, 0/31, 3/1, 3/7, 4/6, 4/10, 5/6, 5/10, 5/16, 6/16, 7/1, 19/1, 19/33, 21/1, 31/24, 31/28, 31/33, 1/13, 1/30, 13/33, 30/8, 30/32, 30/33, 8/32, 8/33, 9/33, 27/23, 27/33, 28/33, 32/14, 32/15, 32/20, 32/22, 32/23, 32/29, 33/15, 33/18, 33/20, 33/22, 33/26, 33/29, 25/23, 25/24, 23/29, 29/26}
{
(\startnode) edge[-,thick,draw opacity=0.75] node {} (\endnode)
};
\node (name) at (0, -2.5) [scale=2.5] {\Large $\psi_2$};
\end{pgfonlayer}
\end{tikzpicture}}\hfill
\resizebox{.2\linewidth}{!}{\begin{tikzpicture}
\foreach \nodename/\x/\y/\cp/\cn/\cw in {
0/1.46/-0.71/0.3640/0.0000/0.6360, 1/1.17/-0.01/0.4131/0.0000/0.5869, 2/-0.11/-0.01/0.2589/0.0000/0.7411, 3/0.68/-0.85/0.4155/0.0000/0.5845, 4/2.18/-1.83/0.0000/0.3054/0.6946, 5/3.21/-1.58/0.0000/0.6261/0.3739, 6/2.87/-1.90/0.0000/0.6261/0.3739, 7/0.93/-0.52/0.4155/0.0000/0.5845, 8/-0.10/0.38/0.0000/0.4038/0.5962, 9/-0.17/1.22/0.0000/0.0265/0.9735, 10/2.61/-1.41/0.0000/0.3054/0.6946, 11/3.27/-0.64/0.1534/0.0000/0.8466, 12/1.05/-1.67/0.1534/0.0000/0.8466, 13/0.29/-0.18/0.1216/0.0000/0.8784, 14/-2.22/1.58/0.0000/0.5067/0.4933, 15/-1.92/1.88/0.0000/0.5081/0.4919, 16/4.00/-2.33/0.0000/0.6729/0.3271, 17/2.36/0.15/0.1534/0.0000/0.8466, 18/-1.43/1.90/0.0000/0.2118/0.7882, 19/0.66/0.42/0.1383/0.0000/0.8617, 20/-2.73/0.85/0.0000/0.5081/0.4919, 21/2.42/-0.28/0.2763/0.0000/0.7237, 22/-0.93/1.84/0.0000/0.5081/0.4919, 23/-2.41/0.35/0.1678/0.0000/0.8322, 24/-2.23/-0.97/1.0000/0.0000/0.0000, 25/-2.65/-0.57/0.8502/0.0000/0.1498, 26/-2.99/1.53/0.0000/0.4146/0.5854, 27/-1.77/-0.17/0.0582/0.0000/0.9418, 28/-1.19/0.11/0.2735/0.0000/0.7265, 29/-2.50/1.11/0.0000/0.4927/0.5073, 30/-0.09/0.83/0.0000/0.3465/0.6535, 31/-1.04/-0.32/0.6585/0.0000/0.3415, 32/-1.43/0.96/0.0000/1.0000/0.0000, 33/-1.25/0.83/0.0000/0.5215/0.4785
}
{
\node (\nodename) at (\x,\y) [shape=circle,inner sep=3pt,draw,thick,fill={rgb:cycle1,\cp;white,\cw;cycle2,\cn}] {}; % cycle2 or cycle5
}
\begin{pgfonlayer}{background}
\path
\foreach \startnode/\endnode in {
2/0, 2/1, 2/3, 2/7, 2/8, 2/9, 2/13, 2/27, 2/28, 2/32, 0/3, 0/4, 0/5, 0/6, 0/7, 0/10, 0/11, 0/12, 0/17, 0/19, 0/21, 0/31, 3/1, 3/7, 4/6, 4/10, 5/6, 5/10, 5/16, 6/16, 7/1, 19/1, 19/33, 21/1, 31/24, 31/28, 31/33, 1/13, 1/30, 13/33, 30/8, 30/32, 30/33, 8/32, 8/33, 9/33, 27/23, 27/33, 28/33, 32/14, 32/15, 32/20, 32/22, 32/23, 32/29, 33/15, 33/18, 33/20, 33/22, 33/26, 33/29, 25/23, 25/24, 23/29, 29/26}
{
(\startnode) edge[-,thick,draw opacity=0.75] node {} (\endnode)
};
\node (name) at (0, -2.5) [scale=2.5] {\Large $\psi_3$};
\end{pgfonlayer}
\end{tikzpicture}}
}
\vspace*{-3mm}
\caption{We remove \textcolor{cycle1}{\bf red} edges from the \textcolor{cycle3}{\bf green} graph to obtain the \textcolor{cycle2}{\bf blue} graph. Eigenvalues interlace (a); eigenvectors $\phi_1,\phi_2,\phi_3$ for \textcolor{cycle3}{\bf green} and $\psi_1,\psi_2,\psi_3$ for \textcolor{cycle2}{\bf blue} highlight common structures.}
%The eigenvectors do not perfecly correspond, \pkw{calling for} the base alignment method of Section~\ref{ssec:basealign}.
% via eigenvalues and eigenvectors could substantially help graph alignment.}
\vspace*{-3mm}
\label{fig:spectra}
\end{figure}
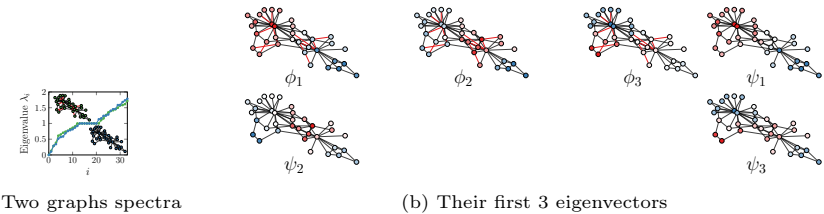

% DM: Don't need to say this. 
%\jh{For this Section, we draw on an approach proposed in~\cite{Kovnatsky13}, which solved an equivalent problem for shapes.} 
\pk{We have hitherto assumed that the graphs to be aligned are isomorphic, hence their eigenvectors correspond to each other with possible sign changes and an orthogonal diagonal mapping matrix $C$ exists. Still, if the graphs are not isomorphic, then their eigenvectors diverge and \pkw{the} diagonal matrix~$C$, which we enforce, cannot capture their relationship well. \dm{Figure~\ref{fig:spectra} highlights this issue: at a high level the eigenvectors underline common structures, but they differ at the node level.} In this case, we need to \emph{align} the two eigenvector bases before we consider aligning corresponding vectors and, eventually, nodes. We express this \dm{\emph{base alignment}~\cite{kovnatsky13coupled}} in terms of \dm{an alignment matrix $M$.} %\pki{An alternative to base alignment is a combined row- and column-wise alignment~\cite{chen2020cone}.}
%, which can also be employed as-is in \thiswork. % why not do it then?
%We delve the investigation of alternative base alignments to future work.
% does rotation fully express base alignment? if so, the above is OK.
%\red{coupling? matrices $F$ and $G$? Provide all missing definitions}.

%\spara{Goal.} eigenvector matrices that are both diagonalizing the graph laplacian (because then they are still approx eigenvectors and have this nice 1 to k property \re{[which property is that? define it.]}) and are coupled (because then they are similar across graphs)

% the title of this paragraph confused WebConf Reviewer.
\dm{\pki{\spara{Diagonalization.}} We align the eigenvectors $\Psi$ by a rotation matrix $M$ so as transform $\Psi$ into $\Phi$: $\hat{\Psi} = \Psi M$. Since $\lapl\Psi = \Psi\Lambda$, finding~$\Psi$ is equivalent to 
 the solution of \pkw{the} quadratic minimization problem $\min_{\Psi} \off(\Psi^\top \lapl_2\Psi)~\text{s.t.}~\Psi^\top\Psi=\eye$ which penalizes elements $\off(\cdot)$ outside of the diagonal, \pkw{in order to preserve orthogonality of the basis.}

Since the eigenvectors are orthonormal, $\Psi^\top\Psi=I$ and for $G_2$'s graph Laplacian eigenvectors $\Lambda_2$, $\Psi^\top \lapl_2\Psi = \Psi^\top \Psi\Lambda_2 = \Lambda_2$, and $M^\top \Psi^\top\lapl_2\Psi M=M^\top\Lambda_2 M$. 
%Here $\Lambda_2$ refers to the eigenvalues of $\lapl_2$. 
Putting the above together, our diagonalizing term is:

\vspace{-2mm}
\[\min_{M} \off(M^\top \Lambda_2M)~\text{s.t.}~M^\top M=I\]
\vspace{-2mm}

As we are minimizing over orthogonal matrices we can equivalently express the objective above as a minimization over orthogonal matrices of size $n \times n$, $S(n,n)$:

\vspace{-2mm}
\[\min_{M\in S(n,n)}\off(M^\top\Lambda_2 M)\]
\vspace{-2mm}

\spara{Coupling.} \pki{In addition,} the correspondence $\tau: G_1\rightarrow G_2$ so that $\phi_i\approx\tau\circ\psi$ translates to 

\vspace{-2mm}
\[\min_\Phi\|F^\top\Phi-G^\top\Psi M\|_F^2\]
\vspace{-2mm}

\noindent where $F$ and $G$ contain each graphs's corresponding functions.
% repetition
%Combining that with the rotation matrix $B$, we get:
%\[\min_{B\in S(k,k)}\mu||F^\top\Phi-G\Psi B||_F^2\]
We combine the minimization terms for diagonalization and coupling, to get the following minimization problem, with regularization factor~$\mu$\dm{\footnote{$\mu =0.132$ in our experiments}}:
% any reason for this value? pk 19oct2020

\vspace{-2mm}
\begin{equation}
\min_{M\in S(n,n)}\off(M^\top\Lambda_2 M)+\mu\|F^\top\Phi-G^\top\Psi M\|_F^2
\label{eq:basealign}
\end{equation}
\vspace{-1mm}

Given that the eigenvectors of isomorphic graphs match with sign changes, we initialize $M$ as a diagonal matrix with $M_{ii}=1$ if   $\|F^\top\phi_i-G^\top\psi_i\|\leq\|F^\top\phi_i+G^\top\psi_i\|$, $M_{ii}=-1$ otherwise.  
% \end{cases}
% \]
}}
% \[M_{i,i}=
% \begin{cases}
% +1 & \text{if }~  \|F^\top\phi_i-G^\top\psi_i\|\leq\|F^\top\phi_i+G^\top\psi_i\|\\
% -1 & \text{otherwise}  
% \end{cases}
% \]
% }}
% \red{What method is used to solve this? What about the text commented out here?}
\dm{Eq.~\eqref{eq:basealign} leads to a \pkw{manifold optimization problem, which we solve by trust-region methods~\cite{absil2007trust}.}}
% as implemented in pymanopt.
% I guess we solve it, do not just report that somebody could solve it! pk 20oct2020

\dm{\spara{Scalability.} \pkw{We} avoid computing all eigenvectors $n\times n$, \pkw{exploiting} the fact that we only need the first $k$ eigenvectors for calculating~$C$ \pkw{(see Section~\ref{sec:matrix_C})}. So we only align the first $k$ eigenvectors of $\Psi$ to the first $k$ eigenvectors of $\Phi$, i.e $\bar{\Phi}=\hat{\Psi} =\bar{\Psi} M$ with $\bar{\Phi}=[\phi_1,\ldots,\phi_k]$ and $\bar{\Psi}=[\psi_1,\ldots,\psi_k]$. Let  $\bar{\Lambda}_2=\diag(\lambda_1,\ldots,\lambda_k)$, the problem in Eq.~\eqref{eq:basealign} becomes

\vspace{-2mm}
\begin{equation}
\min_{M\in S(k,k)}\off(M^\top\bar{\Lambda}_2 M)+\mu\|F^\top\bar{\Phi}-G^\top\bar{\Psi} M\|_F^2
\end{equation}
\vspace{-1mm}

After obtaining $M$, we use the eigenvectors in $\bar{\Phi}$ and the aligned eigenvectors $\hat{\Psi} =\bar{\Psi}M$ in the next step for the final alignment of nodes.}
%\red{What method is used to solve this? What about the text commented out here?}
%We solve Eq.~\ref{eq:basealign} using the trust region method on Riemannian manifolds~\cite{absil2007trust}.
\pkw{Our approach effectively trades off graph alignment with a proxy problem of manifold optimization, which we solve with reasonable accuracy and scalability.}
% comment of R2 refers to Stiefel manifold, which is the set of orthonormal k-frames.
% however, we define M only as orthogonal. 20oct2020

\subsection{Our algorithm: \thiswork}\label{ssec:algorithm}

\pki{Putting it all together, \thiswork  consists of five steps; the pseudocode is given in the supplementary material.} 

\dm{\spara{Steps 1: Compute eigenvectors.} In the first step, calculate the Laplacians $\lapl_1,\lapl_2$ of the two graphs $G_1$ and $G_2$. Then compute the eigenvectors $\Phi,\Psi$ and eigenvalues $\Lambda_1,\Lambda_2$ by the eigendecomposition $\lapl_1=\Phi\Lambda_1\Phi^\top$ and $\lapl_2=\Psi\Lambda_2\Psi^\top$.

\spara{Step 2: Compute corresponding functions.}
In the second step, calculate the matrices of corresponding functions $F=[f_1,\ldots,f_q]$ and $G=[g_1,\ldots,g_q]$ as diagonals of the heat kernel at time steps $[t_1,\ldots,t_q]$ with $f_i= \sum_{j=1}^n e^{-{t_i}\lambda_j}\phi_j \odot \phi_j$ and $g_i$ equivalently using $\Psi$.

\spara{Step 3: Base alignment.}
After  the corresponding functions are calculated, obtain the base alignment matrix $M$ by minimizing Eq.~\ref{eq:basealign}. Then align the first $k$ columns of $\Psi$, denoted by $\bar{\Psi}$ to the corresponding first $k$ columns $\bar{\Phi}$ of $\Phi$ as $\hat{\Psi}=\bar{\Psi}M$.

\spara{Step 4: Calculate mapping matrix.}
Under the assumption that $C$ is a diagonal matrix, calculate its diagonal elements $c_{11},\ldots,c_{kk}$ by solving the least squares problem:
\begin{equation}
\min_{[c_{11},\ldots,c_{kk}]^\top}\left\Vert\begin{bmatrix}
    \diag(g_1^\top\hat{\Psi})\\
    \vdots\\
    \diag(g_q^\top\hat{\Psi})
\end{bmatrix}
    \begin{bmatrix}
    c_{11}\\
    \vdots\\
    c_{kk}
\end{bmatrix}
-  \begin{bmatrix}
    \bar{\Phi}^\top f_1\\
    \vdots\\
    \bar{\Phi}^\top f_q
\end{bmatrix}
\right\Vert_2^2
\label{eq:mapping}
\end{equation}
We then set $C=\diag(c_{11},\ldots,c_{kk})$.

\spara{Step 5: Node alignment.} \pki{To get the final node alignment, we apply a linear assignment algorithm on the rows of $\bar{\Phi}$ and $C^\top\hat{\Psi}$, which hold the indicator function coefficients.}}
%Since the corresponding functions should have the same coefficients on both graphs, we obtain the coefficients of the indicator functions for all nodes as  $\bar{\Phi}^\top$ and $C\hat{\Psi}^\top$. In order to get the final alignment of nodes, align the columns of $\bar{\Phi}^\top$ and $C\hat{\Psi}^\top$ with a linear assignment algorithm.

%\input{figs/alg-grasp.tex} % this belongs here. 20oct2020

%Rev. 4: time complexity in each step of the proposed algorithm is not provided.
\paragraph{Complexity analysis}

\pkw{The computation of the first~$k$ Laplacian eigenvectors
%, used as a base and to compute the corresponding functions,
takes~$\bigO(k\max \{|\E_1|,|\E_2|\})$ by fast methods for diagonally dominant matrices~\cite{koutis2015faster}. Base alignment needs~$\bigO(k^3)$ to solve the orthogonality constraint through trust-region methods. The least-squares method runs in~$\bigO(q\,k)$. The matching step by JV runs in~$\bigO(n^3)$. Overall, the~$\bigO(n^3)$ time factor is dominant.}
%In practice, as we note in Section~\ref{ssec:scalability}, \thiswork runs as fast as REGAL~\cite{heimann18regal} on several datasets with precomputed eigendecomposition.

% differential is the right term, our mistake to call it computational in NetLSD (no idea why)
% This section is an opportunity to show the differences among shape analysis and graphs
\pkw{\paragraph{Connection to Differential Geometry}}

%\jht{
\dmt{\pkw{Our work rests on the theory on~Riemannian manifolds~\cite{gallot1990riemannian} %that studies continuous multidimensional surfaces,
and builds on the analogy between a graph's Laplacian and the continuous Laplace-Beltrami operator~\cite{tsitsulin18kdd}.}}
\section{Experiments}\label{sec:experiments}

% This can be moved in the appendix, if necessary 
% Details of the machines moved in the supplementary material

We experiment on three different real-world networks; \pki{Table~1 in the Supplementary Material lists their properties.} As in~\cite{heimann18regal}, we \pki{randomly permute the node order and} 
%by applying a permutation matrix $P$ to the original adjacency matrix $A_{orig}$, thus $A_{perm}=PA_{orig}P^\top$. 
inject noise by randomly deleting edges with probability $p$. 
%, the latter value being higher than anything used in previous studies~\cite{heimann18regal,nassar2018lowrank}}. % Not true anymore
\pki{We generate a graph for~$p = 0.01$, which we align to 5 noisy graphs, one for each~$p$ value,} ranging from~$0.05$ to~$0.25$; \pki{we measure alignment accuracy as the average ratio of correctly aligned nodes; note that none of the noisy graphs in a pair is a subset of the other.}
%In Section~\ref{ssec:scalability} we show how our method fares in terms of scalability.

\dm{\spara{Baselines.} We compare against the following \pki{established} state-of-the art baselines for \emph{unrestriced} graph alignment.
\begin{itemize}[leftmargin=*]
    \item \textbf{REGAL}~\cite{heimann18regal}: A method based on embeddings \pkw{utilizing} local structural features. %In \pkw{its} original formulation, 
    REGAL %does not perform one-to-one alignment but
    allows \pki{one-to-many matchings}.
    %As the matching process is simply the task of matching the obtained embedding vectors,
    \pki{For the sake of fairness, we let REGAL provide one-to-one matchings using the JV linear assignment algorithm, as \thiswork does; we confirmed that, doing so, it fares better than using nearest neighbors.}
    %to the embeddings.
    \item \textbf{Low Rank EigenAlign (LREA)}~\cite{nassar2018lowrank}: A spectral method that yields one-to-one matchings via the minimization of edge mismatches. %The method outperforms all previous spectral methods like IsoRank~\cite{singh2008global}; hence, it is used as the reference competitor. 
\end{itemize}}
% condensed 20jan2021

\pkw{We eschew a comparison with IsoRank~\cite{singh2008global,liao2009isorankn} and other methods for the alignment of biological networks~\cite{el2015natalie}, since REGAL and LREA significantly outperform those methods.}

\pki{\spara{Parameter tuning.} After experimenting with different settings, we settled for using $k = 20$ eigenvalues and $q = 100$ corresponding functions, which yield best accuracy. We report the tuning experiments in the supplementary material. }

\begin{figure}[h!]
\vspace*{-2mm}
\centering
\begin{tikzpicture}
\begin{groupplot}[group style={
                      group name=myplot,
                      group size= 3 by 1, horizontal sep=0.6cm},height=4cm,width=0.4\linewidth,title style={at={(0.5,0)},anchor=north,yshift=-11mm},ymin=0,ymax=0.8,
                      xmin=0.045,xmax=0.25,xticklabel style={/pgf/number format/fixed},
                      xtick={0.05,0.15,0.25},
                      ymajorgrids,
                      every tick label/.append style={font=\small}
]
\nextgroupplot[
	ylabel={Accuracy},
	xlabel={Noise level},
 	title = {\textbf{Arenas}},
  	legend cell align={left},
 	legend columns=2,
	legend style={at={(1.6,1.1)},anchor=south},
    legend entries={ JV{,} with base alignment, NN{,} with base alignment, JV{,} no base alignment, NN{,} no base alignment,},
    legend style={font=\small}
]
\addplot[color=cycle1,mark=*, mark size=1.5pt] table [x=noise_level, y=jv_ba] {data/ijcai/acc_variants_arenas.csv};
\addplot[color=cycle2!75!white,mark=diamond*, mark size=2pt] table [x=noise_level, y=nn_ba] {data/ijcai/acc_variants_arenas.csv};
\addplot[color=cycle1,dashed,mark=*, mark size=1.5pt] table [x=noise_level, y=jv_no_ba] {data/ijcai/acc_variants_arenas.csv};
\addplot[color=cycle2!75!white,dashed,mark=diamond*, mark size=2pt] table [x=noise_level, y=nn_no_ba] {data/ijcai/acc_variants_arenas.csv};

\nextgroupplot[
	xlabel={Noise level},
 	title = {\textbf{Facebook}},
 	yticklabels={,,},
]
\addplot[color=cycle2!75!white,dashed,mark=diamond*, mark size=2pt] table [x=noise_level, y=nn_no_ba] {data/ijcai/acc_variants_facebook.csv};
\addplot[color=cycle2!75!white,mark=diamond*, mark size=2pt] table [x=noise_level, y=nn_ba] {data/ijcai/acc_variants_facebook.csv};
\addplot[color=cycle1,dashed,mark=*, mark size=1.5pt] table [x=noise_level, y=jv_no_ba] {data/ijcai/acc_variants_facebook.csv};
\addplot[color=cycle1,mark=*, mark size=1.5pt] table [x=noise_level, y=jv_ba] {data/ijcai/acc_variants_facebook.csv};

\nextgroupplot[
	xlabel={Noise level},
 	title = {\textbf{CA-AstroPh}},
 	yticklabels={,,},
]
\addplot[color=cycle2!75!white,dashed,mark=diamond*, mark size=2pt] table [x=noise_level, y=nn_no_ba] {data/ijcai/acc_variants_CA-AstroPh.csv};
\addplot[color=cycle2!75!white,mark=diamond*, mark size=2pt] table [x=noise_level, y=nn_ba] {data/ijcai/acc_variants_CA-AstroPh.csv};
\addplot[color=cycle1,dashed,mark=*, mark size=1.5pt] table [x=noise_level, y=jv_no_ba] {data/ijcai/acc_variants_CA-AstroPh.csv};
\addplot[color=cycle1,mark=*, mark size=1.5pt] table [x=noise_level, y=jv_ba] {data/ijcai/acc_variants_CA-AstroPh.csv};
\end{groupplot}
\end{tikzpicture}
\vspace*{-4mm}
\caption{Accuracy of nearest neighbor and JV matching algorithms.}
%with and without aligning the eigenbases.
\label{fig:variants}
\vspace*{-4mm}
\end{figure}
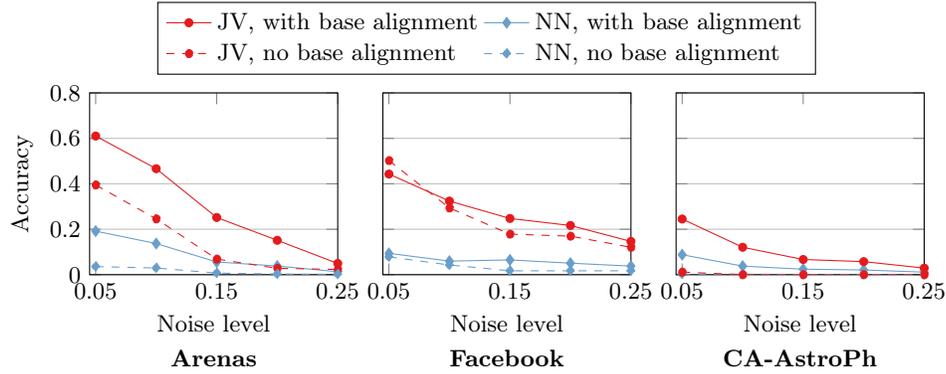

\paragraph{Justifying algorithmic choices}\label{ssec:variants}

\pki{We evaluate the impact of (i) \pki{the choice of} algorithm for node-to-node assignment (Section~\ref{ssec:matching}) and (ii) base alignment (Section~\ref{ssec:basealign}). Figure~\ref{fig:variants} shows that both the JV linear assignment algorithm and base alignment bring a substantial advantage over their rudimentary counterparts, consistently across datasets. 
%The results show improvements from $1.23{\times}$ to $15\times$ when base alignment is used and $1.23{\times}$
In the following experiments, unless otherwise stated, we settle on the variant of \thiswork equipped with base alignment.}

\begin{figure}[h!]
\vspace*{-2mm}
\centering
\begin{tikzpicture}
\begin{groupplot}[group style={
                      group name=myplot,
                      group size= 3 by 1, horizontal sep=0.75cm},height=4cm,width=0.4\linewidth,title style={at={(0.5,0)},anchor=north,yshift=-11mm},ymin=0,ymax=0.8,
                      xmin=0.045,xmax=0.25,xticklabel style={/pgf/number format/fixed},
                      xtick={0.05,0.15,0.25},
                      ymajorgrids,
                      every tick label/.append style={font=\small}
                      ]
\nextgroupplot[
	ylabel={Accuracy},
	xlabel={Noise level},
 	title = {\textbf{Arenas}},
  	legend cell align={left},
 	legend columns=3,
	legend style={at={(1.8,1.1)},anchor=south, font=\small},
    legend entries={\thiswork, REGAL, LREA},
]
\addplot[color=cycle1,mark=*, mark size=1.5pt] table [x=noise_level, y=GrASp] {data/ijcai/acc_other_methods_arenas.csv};
\addplot[color=cycle2,mark=diamond*, mark size=2pt] table [x=noise_level, y=REGAL] {data/ijcai/acc_other_methods_arenas.csv};
\addplot[color=cycle3,mark=square*, mark size=1.5pt] table [x=noise_level, y=LREA] {data/ijcai/acc_other_methods_arenas.csv};

\nextgroupplot[
	xlabel={Noise level},
 	title = {\textbf{Facebook}},
 	yticklabels={,,},
]
\addplot[color=cycle1,mark=*, mark size=1.5pt] table [x=noise_level, y=GrASp] {data/ijcai/acc_other_methods_facebook.csv};
\addplot[color=cycle2,mark=diamond*, mark size=2pt] table [x=noise_level, y=REGAL] {data/ijcai/acc_other_methods_facebook.csv};
\addplot[color=cycle3,mark=square*, mark size=1.5pt] table [x=noise_level, y=LREA] {data/ijcai/acc_other_methods_facebook.csv};

\nextgroupplot[
	xlabel={Noise level},
 	title = {\textbf{CA-AstroPh}},
 	yticklabels={,,},
]
\addplot[color=cycle1,mark=*, mark size=1.5pt] table [x=noise_level, y=GrASp] {data/ijcai/acc_other_methods_CA-AstroPh.csv};
\addplot[color=cycle2,mark=diamond*, mark size=2pt] table [x=noise_level, y=REGAL] {data/ijcai/acc_other_methods_CA-AstroPh.csv};
\addplot[color=cycle3,mark=square*, mark size=1.5pt] table [x=noise_level, y=LREA] {data/ijcai/acc_other_methods_CA-AstroPh.csv};\end{groupplot}
\end{tikzpicture}
\vspace*{-7mm}
\caption{Accuracy compared to REGAL and LREA}\label{fig:quality}
\vspace*{-4mm}
\end{figure}
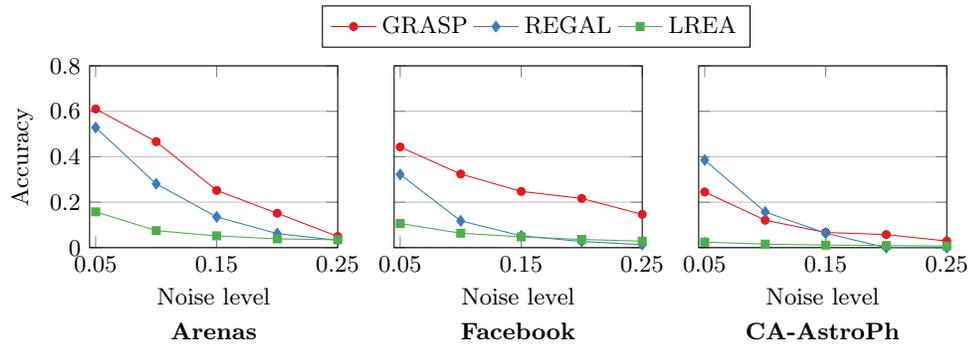

\paragraph{Comparison to previous methods}

%We compare \thiswork to REGAL %~\cite{heimann18regal} and LREA. %~\cite{nassar2018lowrank} 
\pki{Figure~\ref{fig:quality} shows that \thiswork outperforms others by a large margin, achieving $62\%$ accuracy in Arenas \pkw{and} $43\%$ in Facebook with $5\%$ noise,
%\thiswork's behaviour is consistent \pkw{despite} high levels of noise.
and fares at comparably well as REGAL on the CA-AstroPH graph.}
% is this happening?
%\pkw{The accuracy of \thiswork increases more steeply as noise falls.}% Rev. 4

\begin{figure}[h!]
\vspace*{-2mm}
\centering
\begin{tikzpicture}
\begin{groupplot}[group style={
                      group name=myplot,
                      group size= 3 by 1, horizontal sep=0.75cm},height=4cm,width=.4\linewidth,title style={at={(0.5,0)},anchor=north,yshift=-11mm},ymin=0,ymax=1,xticklabel style={/pgf/number format/fixed},ymajorgrids]
\nextgroupplot[
	ylabel={Accuracy},
	xlabel={network \#},
 	title = {\textbf{MultiMagna}},
 	legend cell align={left},
 	legend columns=2,
	legend style={at={(2.85,1.05)},anchor=south east, font=\small},
    legend entries={\thiswork, REGAL},
]
\addplot[color=cycle1,mark=*, mark size=1.5pt] table [x=noise_level, y=GrASp] {data/acc_multi_magna.csv};
\addplot[color=cycle2,mark=diamond*, mark size=2pt] table [x=noise_level, y=REGAL] {data/acc_multi_magna.csv};
\nextgroupplot[
	xlabel={\% of edges},
 	title = {\textbf{Voles}},
 	legend cell align={left},
 	yticklabels={,,},
 	x dir=reverse,
]
\addplot[color=cycle1,mark=*, mark size=1.5pt] table [x=noise_level, y=GrASp] {data/plj.csv};
\addplot[color=cycle2,mark=diamond*, mark size=2pt] table [x=noise_level, y=REGAL] {data/plj.csv};
%\addplot[color=cycle3,mark=square*, mark size=1.5pt] table [x=noise_level, y=LREA] {data/plj.csv};

\nextgroupplot[
	xlabel={\% of edges},
 	title = {\textbf{HighSchool}},
 	yticklabels={,,},
 	x dir=reverse,
]
\addplot[color=cycle1,mark=*, mark size=1.5pt] table [x=noise_level, y=GrASp] {data/prox-high-school.csv};
\addplot[color=cycle2,mark=diamond*, mark size=2pt] table [x=noise_level, y=REGAL] {data/prox-high-school.csv};
%\addplot[color=cycle3,mark=square*, mark size=1.5pt] table [x=noise_level, y=LREA] {data/prox-high-school.csv};
\end{groupplot}
\end{tikzpicture}
\vspace*{-4mm}
\caption{Accuracy compared to REGAL on three real datasets.}
\label{fig:proximity}
%\vspace{-4mm}
\end{figure}
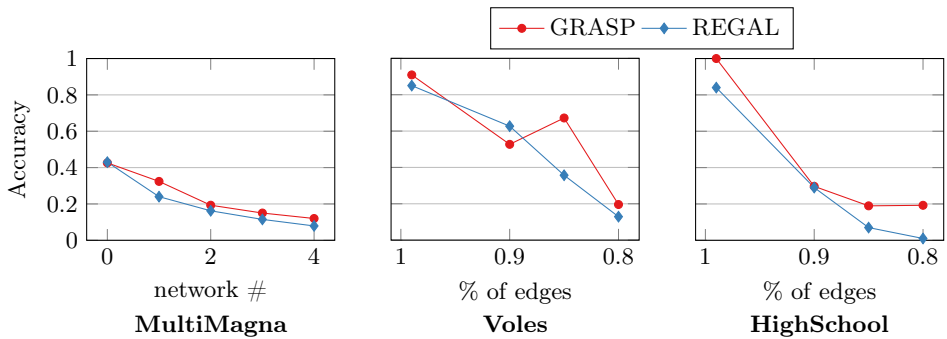

\paragraph{Real world networks}

\pki{%In addition to experiments with artificial noise,
We also try matching among real world networks. MultiMagna %~\cite{vijayan2017multiple} 
is a collection of graphs consisting of a base yeast network and five variations thereof. We match these five variations to the original; Figure~\ref{fig:proximity} presents our results. HighSchool %~\cite{fournet2014contact} 
and Voles %~\cite{davis2015spatial} 
are two evolving proximity networks. We match their latest version to versions at time steps with 80\%, 85\%, 90\%, and 99\% of all edges. Figure~\ref{fig:proximity} shows the results.} \pk{The advantage of \thiswork observed with synthetic noise transfers to real-world alignment problems.}
% condensed 20jan2021

%!TEX root=../main.tex
\begin{figure}[htb]
\vspace*{-2mm}
\centering
\begin{tikzpicture}
\begin{groupplot}[group style={
                      group name=myplot,
                      group size= 2 by 1, horizontal sep=1cm},height=4cm,width=0.49\linewidth,title style={at={(0.5,0)},anchor=north,yshift=-11mm},
                      ymajorgrids,
                      every tick label/.append style={font=\small}
                      ]
\pgfplotstableread{data/time_precomputation.csv}\datatable
\nextgroupplot[
    ylabel={Time, seconts},
    title style={font=\small},
    title = {\textbf{Only alignment}},
    legend cell align={left},
    legend columns=3,
    legend style={at={(1.2,1.1)},anchor=south, font=\small},
    legend entries={\thiswork, REGAL, LREA},
    ybar=0pt,
    bar width=6pt,
    ymode=log,
    ytick={0.01,1,100,10000},
    ymin=0.01,
    ymax=10000,
    enlargelimits=0.15,
    log origin=infty,
    width=0.5\columnwidth,
    height=3.75cm,
    xtick style={draw=none},
    % extra x tick style={
    %     tick label style={rotate=90}},
    xtick=data,
    xticklabel style={rotate=15, font=\small, anchor=base,yshift=-0.3cm, xshift=-0.3cm},
    % every tick label/.append style={font=\tiny},
    xticklabels from table={\datatable}{Graph},
]

\addplot [cycle1, fill=cycle1!50] table [x expr=\coordindex, y={GrASp}]{\datatable};
\addplot [cycle2, fill=cycle2!50] table [x expr=\coordindex, y={REGAL}]{\datatable};
\addplot [cycle3, fill=cycle3!50] table [x expr=\coordindex, y={LREA}]{\datatable};
\nextgroupplot[
    title style={font=\small},
    title = {\textbf{Alignment + precomputation}},
    ybar=0pt,
    bar width=6pt,
    ymode=log,
    ytick={0.01,1,100,10000},
    ymin=0.01,
    ymax=10000,
    enlargelimits=0.15,
    log origin=infty,
    width=0.5\columnwidth,
    height=3.75cm,
    xtick style={draw=none},
    % extra x tick style={
    %     tick label style={rotate=90}},
    xtick=data,
    xticklabel style={rotate=15, font=\small, anchor=base,yshift=-0.3cm, xshift=-0.3cm},
    % every tick label/.append style={font=\tiny},
    xticklabels from table={\datatable}{Graph},
]
\pgfplotstableread{data/time_no_precomputation.csv}\datatable
\addplot [cycle1, fill=cycle1!50] table [x expr=\coordindex, y={GrASp}]{\datatable};
\addplot [cycle2, fill=cycle2!50] table [x expr=\coordindex, y={REGAL}]{\datatable};
\addplot [cycle3, fill=cycle3!50] table [x expr=\coordindex, y={LREA}]{\datatable};
\end{groupplot}
\end{tikzpicture}
%\vspace*{-7mm}
\caption{\pki{Time with offline and online precomputation.}}\label{fig:time}
%\vspace*{-4mm}
\end{figure}
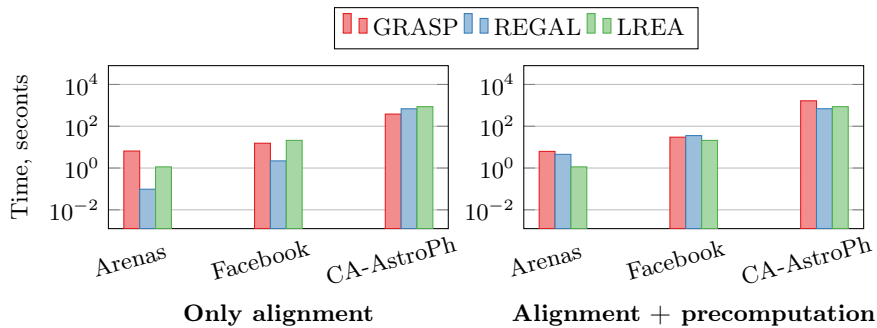

\paragraph{Efficiency}\label{ssec:scalability}

\dm{\pki{Steps 1--3 of \thiswork (Section~\ref{ssec:algorithm}) \pkw{can be performed offline,} while REGAL also allows for} precomputation of representations. Figure~\ref{fig:time} (left) shows the time to compute the alignments \pki{after precomputation}. \thiswork outperforms REGAL and LREA in the largest CA-AstroPh data.
%and performs, in the other cases as well as the optimized low-rank method of LREA.
Figure~\ref{fig:time} (right) shows the time \pki{with online precomputation}; REGAL does not exhibit any substantial advantage even in the smaller Arenas and Facebook graphs, while \thiswork attains more accurate results with a negligible increase in time.} \pki{We obtained similar results on real-world network matching tasks.}
% condensed 20jan2021

%\newpage
% \begin{figure}
%     \centering
%     \includegraphics[scale=0.5]{figs/contacts-prox-high-school-2013.png}
%     \caption{contacts-prox-high-school-2013, full graph matched to 80\%, 90\%, 95\% and 99\%. of the graph (edges) considering time}
%     \label{fig:bio}
% \end{figure}

%\cmt{Prove that we are not that slow afterall} 

%\input{content/exp_figures3.tex}

%\begin{table}[h!] %not correct values
%\begin{center}
%\begin{tabular}{| l || l | l | l|}
%\hline
%                   & Arenas Email & facebook-ego & bingham82\\ \hline\hline
%GASP (jv\!+\!icp)  &   4.15       &     32.29    &  186.19\\ \hline
%GASP (jv)          &  0.83        &    17.75     &  75.51\\ \hline
%REGAL              &  0.21        &    1.91      & 75.67\\ \hline
%LREA               &              &              &      \\ \hline
%\end{tabular}
%\vspace{-3mm}
%\caption{Time with precomputation}\label{tab:time}
%\vspace{-3mm}
%\end{center}
%\end{table}

%\subsection{Parameters}
%\subsection{Variants}
%\subsection{Others}
%\subsection{Running Time}
%!TEX root=../main.tex
\section{Conclusion}\label{sec:conclusion}

\pki{We proposed a graph alignment method using their Laplacian eigenvectors. \pkw{We establish} a functional correspondence among pre-aligned eigenvectors, capturing \pkw{multiscale graph properties} and \pkw{extract a linear assignment} among matrix columns, attaining superior \pkw{alignment quality} over the state of the art.
%\pkw{across high noise levels and real-world graphs.}
%, with noise levels higher than anything used in previous studies}. 
\pki{To our knowledge, this is the first work to apply a functional alignment primitive to graph alignment.} %\cmt{real?}. \cmt{Moreover, by approximating the first eigenvectors, our method scales to large graphs of X nodes.} 
In the future, we plan to extend our method \pkw{to} partial correspondences}
%\pkw{towards} %flexible definitions of subgraph isomorphism\pkw{, including 
%the case of matching graphs with unequal numbers of nodes}
\pki{and examine to what extent our representations can be employed within the framework of~\cite{chen2020cone}.}
% remark from Sec2 comes here.

%\scriptsize
\small
\bibliographystyle{splncs04}
\bibliography{bibliography}

\end{document}

\pagestyle{plain} % temporary (can be in submission)

\maketitle

\begin{table*}[t!]
% \vspace{-3mm}
% \begin{center}
\newcolumntype{C}{>{\centering\arraybackslash}X}
\begin{tabularx}{\linewidth}{ lrrXX}
\toprule
\textbf{Dataset} & $|V|$ & $|E|$ & \textbf{Network type} & \textbf{Alignments}\\ \midrule
Arenas Email \cite{kunegis2013konect} & 1\,133  & 5\,451   & communication & generated\\
Facebook-ego \cite{snapnets}          & 4\,039  & 88\,234  & social & generated\\ 
CA-AstroPh \cite{snapnets}                   & 17\,903 & 197\,031 & collaboration & generated\\
MultiMagna \cite{vijayan2017multiple} & 1\,004 & 8\,323 & biological & provided\\
HighSchool~\cite{fournet2014contact} &  327 & 5\,818 & proximity & provided\\
Voles~\cite{davis2015spatial} &  712 & 2\,391 & proximity & provided\\

\bottomrule
\end{tabularx}
\caption{Datasets used in our evaluation, $|V|$ number of nodes, $|E|$ number of vertices.}\label{tbl:datasets}
% \end{center}
\vspace{-3mm}
\end{table*}

\begin{figure*}
\centering
\begin{tikzpicture}
\begin{groupplot}[group style={
                      group name=myplot,
                      group size= 4 by 1, horizontal sep=.5cm},height=4cm,width=0.3\linewidth,title style={at={(0.5,0)},anchor=north,yshift=-11mm},ymin=0,ymax=0.8,xmin=10,xmax=100]
\nextgroupplot[
	ylabel={Accuracy},
	xlabel={$k$},
 	title = {Arenas, $q=100$},
 	legend columns=4,
	legend style={at={(3,1.05)},anchor=south east},
    legend entries={5\% noise, 10\% noise, 15\% noise},
]
\addplot[color=cycle1!50!white,mark=*, mark size=1.5pt] table [x=k, y=noise_5] {data/ijcai/acc_params_k_arenas.csv};
\addplot[color=cycle1!75!white,mark=square*, mark size=1.5pt] table [x=k, y=noise_10] {data/ijcai/acc_params_k_arenas.csv};
\addplot[color=cycle1,mark=diamond*, mark size=2.5pt] table [x=k, y=noise_15] {data/ijcai/acc_params_k_arenas.csv};

\nextgroupplot[
	xlabel={$k$},
 	title = {Facebook, $q=100$},
 	yticklabels={,,},
]
\addplot[color=cycle1!50!white,mark=*, mark size=1.5pt] table [x=k, y=noise_5] {data/ijcai/acc_params_k_facebook.csv};
\addplot[color=cycle1!75!white,mark=square*, mark size=1.5pt] table [x=k, y=noise_10] {data/ijcai/acc_params_k_facebook.csv};
\addplot[color=cycle1,mark=diamond*, mark size=2.5pt] table [x=k, y=noise_15] {data/ijcai/acc_params_k_facebook.csv};

\nextgroupplot[
	xlabel={$q$},
 	title = {Arenas, $k=20$},
 	yticklabels={,,},
]
\addplot[color=cycle1!50!white,mark=*, mark size=1.5pt] table [x=q, y=noise_5] {data/ijcai/acc_params_q_arenas.csv};
\addplot[color=cycle1!75!white,mark=square*, mark size=1.5pt] table [x=q, y=noise_10] {data/ijcai/acc_params_q_arenas.csv};
\addplot[color=cycle1,mark=diamond*, mark size=2.5pt] table [x=q, y=noise_15] {data/ijcai/acc_params_q_arenas.csv};

\nextgroupplot[
	xlabel={$q$},
 	title = {Facebook, $k=20$},
 	yticklabels={,,},
]
\addplot[color=cycle1!50!white,mark=*, mark size=1.5pt] table [x=q, y=noise_5] {data/ijcai/acc_params_q_facebook.csv};
\addplot[color=cycle1!75!white,mark=square*, mark size=1.5pt] table [x=q, y=noise_10] {data/ijcai/acc_params_q_facebook.csv};
\addplot[color=cycle1,mark=diamond*, mark size=2.5pt] table [x=q, y=noise_15] {data/ijcai/acc_params_q_facebook.csv};
\end{groupplot}
\end{tikzpicture}
\vspace*{-7mm}
\caption{Accuracy on the facebook and arenas graphs for different numbers of corresponding functions $q$ and different numbers of eigenvectors $k$ for noise levels 5\%, 10\% and 15\%.}
\label{fig:parameters}
\end{figure*}

% \appendix
We highlight details and experiments for reproducibility. We answer the following questions from the subsmission system. 

\question{\textbf{Theoretical claims} -- Does this paper make theoretical claims?}

\answer{\emph{All assumptions and restrictions are stated clearly and formally: } We motivate in Section~3 and Section~4 our choice of functional alignment, the basis (Section~4.1) and the function (Section~4.2). We also report the theoretical connections with differential geometry in Section~4.6.}

\question{\textbf{Data sets} -- Does this paper rely on one or more data sets?}

\answer{\emph{All novel datasets introduced in this paper are described in detail in a data appendix, including the collecting procedure and data statistics --
All datasets drawn from the existing literature (potentially including authors’ previously published work) are accompanied with appropriate citations:}
 We perform experiments on six real datasets, three of which (Arenas Email, Facebook-ego, CA-AstroPh) having alignments \emph{generated} from random deletion of edges on the original graph, and three of them (MultiMagna, HighSchool, Voles) with provided alignments. The characteristics of the datasets and the literature from which each dataset can be downloaded are described in Table~\ref{tbl:datasets}.

\emph{The preprocessing details of datasets are included:} The preprocessing is described in Section~5 (first paragraph). The use of the datasets with \emph{provided alignment} is described in Section~5 (paragraph ``Real World Networks'').}

\question{\textbf{Models and Algorithms} -- Does this paper include novel models and algorithms?} 

\answer{\emph{The input and output, mathematical setting, and algorithms are clearly introduced:} Section~3 introduces the preliminaries. the background notation, and the problem.  The pseudocode of \thiswork is shown in Algorithm~\ref{alg:grasp}. The algorithm's steps are described in Section~4.6. 

\begin{algorithm}[!h]
\begin{algorithmic}[1]
% \small
\Require{Graphs $G_1 = (V_1, E_1)$, $G_2 = (V_2, E_2)$}
\Params{Eigenvectors $k$, functions $q$, times $t=[t_1,\ldots,t_q]$}
%\Ensure{An action $(v, \fidelity)$}
\smallComment{Step 1: Eigendecomposition of $\lapl$}
\State $\lapl_1\gets\eye-D_1^{-\frac{1}{2}}A_1D_1^{-\frac{1}{2}}$; $\lapl_2\gets\eye-D_2^{-\frac{1}{2}}A_2D_2^{-\frac{1}{2}}$
% \smallComment{Step 2: Eigendecomposition of Laplacians}
\State $\Phi,\Lambda_1 \gets \eig(\lapl_1)$; $\Psi,\Lambda_2 \gets \eig(\lapl_2)$
\smallComment{Step 2: Corresponding functions}
\ForAll{$t_i$ in $t$} 
\State $f_i\gets \sum_{j=1}^n e^{-{t_i}\lambda_j}\phi_j \odot \phi_j$
\State $g_i\gets \sum_{j=1}^n e^{-{t_i}\lambda_j}\psi_j \odot \psi_j$
\EndFor
\State $F \gets [f_1, \ldots, f_q]$; $G \gets [g_1, \ldots, g_q]$
\smallComment{Step 3: Base alignment}
\State $\bar{\Phi}=[\phi_1,\ldots,\phi_k]$; $\bar{\Psi}=[\psi_1,\ldots,\psi_k]$
\State $M\gets\min_{M\in S(k,k)}\off(M^\top\bar{\Lambda}_2 M)+\mu\|F^\top\bar{\Phi}-G\bar{\Psi} M\|_F^2$
\State $\hat{\Psi}=\bar{\Psi}M$
\smallComment{Step 4: Mapping matrix}
\State \dm{$[c_{11},\ldots,c_{kk}]\gets$  Least squares solution of Eq.~5.}
% \begin{equation*}
% \min_{[c_{11},\ldots,c_{kk}]^\top}\left\Vert\begin{bmatrix}
%     \diag(g_1^\top\hat{\Psi})\\
%     \vdots\\
%     \diag(g_q^\top\hat{\Psi})
% \end{bmatrix}
%     \begin{bmatrix}
%     c_{11}\\
%     \vdots\\
%     c_{kk}
% \end{bmatrix}
% -  \begin{bmatrix}
%     \bar{\Phi}^\top f_1\\
%     \vdots\\
%     \bar{\Phi}^\top f_q
% \end{bmatrix}
% \right\Vert_2^2
% \end{equation*}
\State $C\gets\diag([c_{11},\ldots,c_{kk}]^\top)$
%\smallComment{align eigenvectors with C}
\smallComment{Step 5: Matching by assignment}
\State $N\gets$ Matching of columns of $\bar{\Phi}^\top$ to those of $C\hat{\Psi}^\top$ 
\State\Return $N$
\end{algorithmic}
\caption{\thiswork}\label{alg:grasp}
\end{algorithm}

\emph{The complexity analysis including time, space, and sample size is presented:} The complexity analysis is presented in Section~4.6 (paragraph ``Complexity Analysis''). The space complexity is $\bigO(n\max\{q,k\})$ to store the eigenvectors.} 

\question{\textbf{Experimental Results} -- Does this paper include computational experiments?}

\answer{\emph{The evaluation metrics are formally described and the reasons for choosing these metrics are explained:} We measure alignment accuracy as the average ratio of correctly aligned nodes. Accuracy has been used in all previous studies to measure alignment quality. We also report time in Section~5 (paragraph ``Efficiency'')

\emph{Detailed hyperparameter settings of algorithms are described, including the ranges and how to select the best ones:} We report tuning experiments for the algorithm's parameters $q$ and $k$ and the pseudocode of algoorithms.

\para{Varying the number of eigenvectors $k$.}
\dm{The number of eigenvectors affects the quality of the alignment as different eigenvectors potentially capture structures at a different scale. The first two charts on the left in Figure~\ref{fig:parameters} show the accuracy of \thiswork as a function of the number of eigenvectors at 5\% (top line), 10\% (middle line), 15\% (bottom line) noise level when the number of corresponding functions $q=100$. 
The first few eigenvectors with $k = 20$ improve the quality of the alignment, which gradually decreases if we consider more eigenvectors. Such behaviour is expected, as eigenvectors that correspond to larger eigenvalues are representative of medium- to small- scale structures and exhibits large noise. Interestingly, the first eigenvectors can also be computed efficiently with iterative methods.} \pkw{We set $k = 20$ as the default value in our experiments.}

\para{Varying the number of corresponding functions $q$.} 
\dm{
The number of corresponding functions influences the granularity at which the heat kernel is linearly sampled over the time interval $[0.1, 50]$. 
More corresponding functions lead to a more precise representation of the graph at a cost of computation time. 
Figure~\ref{fig:parameters} (the two charts on the right) show the relationships between quality and corresponding functions. 
In both datasets, the quality \jh{slightly increases with $q$ at different noise levels. We set $q = 100$} in our experiments.}

\emph{The computing infrastructure for running experiments is described in detail, including GPU/CPU models, amount of memory, operating system:} We run the experiments on a 40-core  Intel Xeon CPU E5-2687Wv3, 3.10GHz machine with 368Gb RAM on python 3.6.9 and Ubuntu Linux~4.15.0. No GPU acceleration is used in the experiments. We use pymanopt~\cite{townsend2016pymanopt} for solving the base alignment problem in Eq.~4.
}
% \emph{Analysis of experiments goes beyond single-dimensional summaries of performance (e.g., average; median) to include measures of variation, confidence, or other distributional information.}
% }

\question{\textbf{Code}- Please also complete the list of source code below.}

\answer{\emph{All source code required for conducting experiments is uploaded as attachments.} The source code is submitted alongside with the paper in a zip file. 

\emph{All source code required for conducting experiments will be made publicly available upon publication of the paper with a license that allows free usage for research purposes:} We will release our code on Github with an MIT license for unrestricted usage. We will also ensure the code is well-documented with a README.

\emph{A README is provided to explain the structure of the source code, and how to obtain different results in the submission with corresponding scripts.} A README file is attached to the code.
}
\bibliographystyle{named}
\bibliography{bibliography}